\documentclass[fleqn,usenatbib]{mnras}
\pdfoutput=1

\usepackage[T1]{fontenc}
\usepackage{ae,aecompl}
\usepackage{color, soul}
\usepackage{verbatim}
\usepackage{units}
\usepackage{subfig}
\usepackage{pdflscape}
\usepackage{svg}
\usepackage{bm}
\usepackage{stfloats}
\usepackage{float}
\usepackage{multicol}
\usepackage{hyperref}
\usepackage{multirow}

\usepackage{graphicx}	
\usepackage{amsmath}	
\usepackage{amssymb}	
\usepackage{epstopdf}
\usepackage[normalem]{ulem} 

\defcitealias{Lu2013}{L13}
\defcitealias{Bartko2010}{B10}
\defcitealias{Salpeter1955}{S55}
\defcitealias{Evans2022}{E22}

\newcommand\clearrow{\global\let\rowmac\relax}
\newcommand{\cmmnt}[1]{\ignorespaces}

\title[(Evolved) \textit{Gaia} HVSs and the GC]{Constraints on the Galactic Centre environment from \textit{Gaia} hypervelocity stars II: The evolved population}

\author[Evans et al.]{
F. A. Evans$^{1}$\thanks{E-mail: evans@strw.leidenuniv.nl},
T. Marchetti$^{2}$,
E. M. Rossi$^{1}$ \\
$^{1}$Leiden Observatory, Leiden University, PO Box 9513, NL-2300 RA Leiden, The Netherlands\\
$^{2}$European Southern Observatory, Karl-Schwarzschild-Strasse 2, 85748 Garching bei M{\"u}nchen, Germany \\
}

\date{Accepted XXX. Received YYY; in original form ZZZ}

\pubyear{2021}

\begin{document}
\label{firstpage}
\pagerange{\pageref{firstpage}--\pageref{lastpage}}
\maketitle

\begin{abstract}
A dynamical encounter between a stellar binary and Sgr A* in the Galactic Centre (GC) can tidally separate the binary and eject one member with a velocity beyond the escape speed of the Milky Way. These hypervelocity stars (HVSs) can offer insight into the stellar population\textcolor{black}{s in the GC environment.} In a previous work, our simulations showed that the lack of main sequence HVS candidates \textcolor{black}{with precise astrometric uncertainties and radial velocities in} current data releases from the \textit{Gaia} space mission  places a robust upper limit on the ejection rate of HVSs from the GC of $3\times10^{-2} \, \mathrm{yr^{-1}}$. We improve this constraint in this work by additionally considering the absence of post main sequence HVSs in \textit{Gaia} Early Data Release 3 as well the existence of the HVS candidate S5-HVS1. This evidence offers degenerate joint constraints on the HVS ejection rate and the stellar initial mass function (IMF) in the GC. For a top-heavy GC IMF as suggested by recent works, our modelling motivates an HVS ejection rate of $\eta=0.7_{-0.5}^{+1.5} \times10^{-4} \, \mathrm{yr^{-1}}$. This preferred ejection rate can be as large as $10^{-2} \, \mathrm{yr^{-1}}$ for a very top-light IMF and as low as 10$^{-4.5} \, \mathrm{yr^{-1}}$ if the IMF is extremely top-heavy. Constraints will improve further with future \textit{Gaia} data releases, regardless of how many HVS candidates are found therewithin. 
\end{abstract}


\section{Introduction}

The pursuit of identifying and studying fast-moving Milky Way stars is now entering its second century \citep{Barnard1916,vanMaanen1917,Oort1926}. With peculiar velocities of tens to hundreds of $\mathrm{km \ s^{-1}}$, so-called runaway stars have long been recognized for their potential to provide insight into the dynamical and astrophysical phenomena responsible for accelerating them, particularly the disruption of binaries following a supernova explosion \citep{Blaauw1961,Boersma1961, Tauris1998, Eldridge2011, Renzo2019} and ejections following dynamical encounters in young stellar systems \citep{Poveda1967, Leonard1991, Perets2012, Oh2016}.

While very fast ejections from each of the above mechanisms are possible in rare circumstances \citep{Leonard1990, Portegies2000, Gvaramadze2009VMS, Tauris2015, Evans2020}, for early-type stars the upper limit on the ejection velocity appears to be $\sim$450 $\mathrm{km \ s^{-1}}$ \citep[see][]{Irrgang2019}. Alternative mechanisms must therefore be invoked for stars with peculiar velocities above the Galactic escape velocity \citep[$\sim500 \mathrm{km \ s^{-1}}$ at the Solar position,][]{Deason2019, Koppelman2021, Necib2022}. Arguably the most attractive option of these alternatives is the \citet{Hills1988} mechanism, in which a stellar binary is disrupted following an dynamical encounter with Sgr A*, the $\sim$4$\times10^{6} \, \mathrm{M_\odot}$ supermassive black hole located in the Galactic Centre \citep[GC;][]{Ghez2008, Genzel2010}. One member of the former binary is ejected as a \textit{hypervelocity star} (HVS) at a characteristic velocity \textcolor{black}{up to and beyond} $\sim$1000 $\mathrm{km \ s^{-1}}$ \citep{Hills1988, Gould2003, Yu2003, Bromley2006, Generozov2021}. On their outward journeys through the Galaxy and beyond to intergalactic space, a sizeable sample of these stars would serve as a valuable dynamical tracer for the Galactic potential out to large distances \citep{Gnedin2005, Yu2007, Kenyon2008, Kenyon2014, Contigiani2019}.

This Hills mechanism is also an enticing explanation for the S-star cluster, a population of early-type stars on close, eccentric orbits \citep{Gillessen2009, Gillessen2017} about Sgr A* within the innermost arcsecond ($\sim$0.04 pc) of the Galaxy, where tidal forces from Sgr A* are thought to impede in-situ star formation \citep[but see][]{Habibi2017}. The scattering of stellar binaries from the nuclear star cluster \citep[see][]{Launhardt2002, Schodel2014, schodel2014a&a, Neumayer2020} or from substructures within it onto orbits \textcolor{black}{with close periapses to} Sgr A* leads to Hills exchange encounters -- one star is ejected as an HVS and its former companion remains bound to Sgr A* as an S-star \citep{Perets2007, Madigan2009, Zhang2013, Madigan2014, Generozov2020}. Since they are GC-born objects located elsewhere on the sky, a sizeable, uncontaminated sample of HVSs with precisely-known kinematics would also be useful as a tool to study the stellar environment in the GC \citep{Rossi2017, Evans2022}, where direct observation is complicated by extreme and highly inhomogenous dust extinction \citep[see][for a review]{Schodel2014}. 

After the first serendipitous HVS detections \citep{Brown2005, Edelmann2005, Hirsch2005}, several dozen HVS candidates were reported in the decade following \citep[e.g.][]{Brown2006, Heber2008, Tillich2009, Irrgang2010, Brown2012, Brown2014, Zhong2014, Palladino2014}. See \citet{Brown2015rev} for a review on HVSs. Of particular note is S5-HVS1 \citep{Koposov2020}, a $2.35 \, \mathrm{M_\odot}$ HVS candidate discovered in the Southern Stellar Stream Spectroscopic Survey \citep[S$^5$;][]{Li2019}. In contrast to other HVS candidates, the trajectory of S5-HVS1 unambiguously implies an origin in the GC.

Recently, our knowledge on the kinematics of Milky Way stars both fast \textit{and} slow has been revolutionized by the European Space Agency's ongoing \textit{Gaia} mission \citep{Gaia2016,Gaia2018,Gaia2021, Gaia2022DR3}. With unprecedented astrometric measurements of $\sim$2 billion Galactic sources and radial velocity measurements for a subset of $\sim$tens of millions of cool, bright stars, \textit{Gaia} has demystified the origins of some HVS candidates \citep{Irrgang2018, Brown2018, Erkal2019, Kreuzer2020} recategorized others as spurious detections and/or bound stars \citep{Boubert2018, Boubert2019, Marchetti2021}, and discovered new (candidate) stars with extreme velocities \citep{Bromley2018, Shen2018, Hattori2018, Du2019, Luna2019, Marchetti2019, Li2021, Marchetti2021}. While these unbound star candidates are each fascinating in their own right, it is conspicuous that promising genuine HVS candidates, i.e. unbound stars with precisely known kinematics and trajectories that uncontroversially suggest an origin in the GC,  have yet to be unearthed using data solely available in the \textcolor{black}{radial velocity catalogues of \textit{Gaia} Data Release 1, Data Release 2 (DR2) nor Early Data Release 3 (EDR3).} 

\textcolor{black}{Though somewhat disheartening, \citet{Kollmeier2009, Kollmeier2010} showed how the absence of confident HVS candidates in a particular survey is in itself a valuable observational result}. The Galactic HVS population is directly related to the stellar environment in the GC \citep{Sari2010, Kobayashi2012, Rossi2014}. If the GC were particularly effective at ejecting HVSs, \textit{Gaia} should be able to see them. An absence of detected HVSs in \textit{Gaia} DR2/EDR3 would then refute models of the inner parsecs of the Galaxy incompatible with this null detection. We investigated this in \citet{Evans2022}, hereafter \citetalias{Evans2022}. Simulating the ejection of main sequence HVSs from the GC, we showed that the lack of high-confidence HVS candidates in \textit{Gaia} DR2/EDR3 dictates that HVSs must be ejected from the GC at a rate no larger than $\sim3\times10^{-2} \, \mathrm{yr^{-1}}$. Forecasting ahead, we showed the HVS populations (or lack thereof) to appear in the then future with the release of \textit{Gaia} Data Release 3 (DR3) and Data Release 4 (DR4) would improve this constraint considerably and would additionally constrain the slope of the stellar initial mass function (IMF) in the GC. We showed as well that while the population of \textit{Gaia}-visible HVSs \textit{does} depend on the orbital separation and mass ratio distributions among the HVS progenitor binaries, this dependence is too subtle to provide meaningful constraints on these properties given the current paucity of positive detections. 

In this work, we expand upon \citetalias{Evans2022} in several ways. We simulate the ejection of HVSs which have evolved off the main sequence, either before or after their ejection from the GC. While post-main sequence HVSs are a minority of Galactic HVSs, they offer a number of advantages in the particular context of the \textit{Gaia} DR2/EDR3 radial velocity catalogues. For the \textit{Gaia} DR2/EDR3 spectroscopic pipeline to assign a validated radial velocity to a source, it must be brighter than the 12th magnitude in the \textit{Gaia} $G_{\rm RVS}$ band \citep{Gaia2018} and it must have an effective temperature in the range $3500 \, \mathrm{K} < T_{\rm eff} < 6900 \, \mathrm{K}$ \citep{Katz2019}. This temperature condition is restrictive for main sequence HVSs, as the hot-end limit corresponds roughly to a stellar mass of $1.5 \, \mathrm{M_\odot}$ \citep{Pecaut2013}. A main sequence star of this mass must be less than $\sim$1 kpc away to satisfy $G_{\rm RVS}<12$. Since HVSs are ejected isotropically from the GC $\simeq8 \, \mathrm{kpc}$ away from Earth, only a slim minority of HVSs are closer than $1 \, \mathrm{kpc}$ away \citepalias[][c.f. fig. A2]{Evans2022}. Post-main sequence stars, however, are significantly cooler and intrinsically much brighter than main sequence stars at fixed stellar mass. Stellar evolution models predict that nearly \textit{all} giants and supergiants are cooler than $T_{\rm eff} = 6900 \, \mathrm{K}$. With these cooler temperatures and higher luminosities, post-main sequence HVSs up to $\sim10 \, \mathrm{kpc}$ away can appear in the \textit{Gaia} DR2/EDR3 radial velocity catalogue. Despite comprising only 8 per cent of total Galactic HVSs, the effective observation volume for post-main sequence HVSs is 1000 times larger than for main sequence HVSs. When combined, the more-numerous main sequence HVSs and the easier-to-detect evolved HVSs allow stricter constraints on the GC stellar environment than the main sequence HVSs alone. 

To keep focus on \textit{Gaia}, the only existing HVS observational evidence we considered in \citetalias{Evans2022} was the lack of HVS candidates with radial velocities in DR2/EDR3. With the groundwork laid, however, more evidence can be considered. In particular we consider the existence of S5-HVS1 as well. While it has \textit{Gaia} astrometry, S5-HVS1 is neither bright nor cool enough to appear in the \textit{Gaia} DR2/EDR3 radial velocity catalogues. Even so, it remains to date the only uncontroversial HVS candidate. A robust model of the GC stellar environment and the ejection of HVSs must make predictions simultaneously consistent with the lack of HVS candidates in \textit{Gaia} DR2/EDR3 and the existence of S5-HVS1.

In Sec. \ref{sec:methods} of this work we describe our HVS ejection model, in which we eject mock populations of both main sequence and evolved HVSs from the GC, propagate them through the Galaxy and obtain mock observations. In Sec. \ref{sec:results} we present our results, exploring the population of HVSs at different evolutionary stages we predict to lurk in current and future \textit{Gaia} data releases. We use these predictions to investigate how the lack of HVSs in \textit{Gaia} EDR3 and the existence of S5-HVS1 constrains the GC stellar environment, and show how these constraints will improve with future \textit{Gaia} data releases. In Sec. \ref{sec:discussion} we discuss the implications of these results and their caveats, and highlight interesting sub-populations within our mock HVS populations. Finally, we present our conclusions in Sec. \ref{sec:conclusions}.

\section{Ejection Model} \label{sec:methods}

Our model for the generation, ejection, propagation and observation of our mock HVS populations is similar to the Monte Carlo (MC) model used in \citetalias{Evans2022}. We briefly describe the model and its updates here and refer the reader to \citetalias{Evans2022} \citep[see also][]{Marchetti2018, Evans2021} for more detailed information. \textcolor{black}{The model we describe in this Section is implemented in the publicly available \texttt{PYTHON} package \texttt{speedystar}\footnote{\url{https://github.com/fraserevans/speedystar}}.}

\subsection{Generating and ejecting HVSs}

In our HVS ejection model, \textcolor{black}{four parameters define the initial conditions of the HVS progenitor binary: the zero-age main sequence (ZAMS) mass of the larger star in the binary ($m_{\rm p}$), the mass ratio $q\equiv m_{\rm s}/m_{\rm p}$ between the less-massive secondary ($m_{\rm s}$) and primary; $a$, the orbital semi-major axis of the binary at the moment it is tidally separated; and $\xi \equiv \mathrm{log_{10}}[Z/Z_\odot]$, the total stellar metallicity for both stars in the binary. We draw} $m_{\rm p}$ in the range [0.1, 100] $\mathrm{M_\odot}$ assuming a  single power-law IMF with slope $\kappa$, i.e. $f(m_{\rm p})\propto m_{\rm p}^{\kappa}$. \textcolor{black}{We assume binary mass ratios are also distributed as a power law with log-slope $\gamma$, confined to the range $0.1\leq q \leq 1 $.} Orbital semi-major axes are drawn assuming binary orbital periods are distributed as $f(\mathrm{log}P)\propto (\mathrm{log}P)^{\pi}$, where $\pi$ is the log-period power law slope. Minimum periods are set following the approximations of \citet{Eggleton1983} to ensure that neither member star of the binary is filling its Roche lobe at the moment of tidal separation. With this minimum period set, interaction between the stars is minimal and each can be assumed to evolve as if it were isolated. \textcolor{black}{Finally, for each system we sample $\xi$ in the range [-0.25, +0.25] assuming a solar metallicity of $Z_{\odot}=0.0142$ \citep{Asplund2009}. Stellar metallicities in the GC environment exhibit a significant spread with a slightly super-solar mean \citep{Do2015, FeldmeierKrause2017, Rich2017, FeldmeierKrause2020, Schodel2020}.}

\textcolor{black}{There are numerous indications that} the IMF in the GC, at least among certain stellar substructures, is top-heavy \citep[e.g.][]{Paumgard2006, Maness2007, vonFellenberg2022}. We choose $\kappa=-1.7$ as our fiducial power law IMF slope following \citet{Lu2013}, hereafter \citetalias{Lu2013}, based on Keck observations of \textcolor{black}{the young stellar population within the innermost half-parsec of the GC.} We will also at times highlight a model with a particularly top-heavy IMF in which $\kappa=-0.45$ \citep[hereafter \citetalias{Bartko2010}]{Bartko2010} and one which follows a canonical \citet{Salpeter1955} (hereafter \citetalias{Salpeter1955}) IMF, i.e. $\kappa=-2.35$. In \citetalias{Evans2022} we showed that the number of high-confidence HVS candidates appearing in current and future \textit{Gaia} data releases is not particularly sensitive to the binary mass ratio distribution log-slope $\gamma$ and the log-period power law slope $\pi$. We have confirmed that this remains true for evolved HVSs. When generating mock HVS populations we therefore sample $\gamma$ and $\pi$ uniformly in the ranges [-3,+2] and [-2,+2] respectively, capturing the range of values reported in studies of massive binaries in star-forming regions in the Galaxy and the Magellanic Clouds \citep{Sana2012,Sana2013,Moe2013,Dunstall2015,Moe2015}.

Following the tidal separation of the binary, one star is ejected while the other remains bound to Sgr A*. \textcolor{black}{If the binary approached Sgr A* on a parabolic orbit, the primary and secondary members of the binary are equally likely to be ejected as the HVS} \citep{Sari2010,Kobayashi2012}. We therefore randomly designate one star as the ejected one. It has a stellar mass $m_{\rm ej}$ and its ejection velocity is calculated analytically \citep{Sari2010, Kobayashi2012,Rossi2014}:
\begin{equation}
    v_{\rm ej} = \sqrt{\frac{2Gm_{\rm c}}{a}} \left( \frac{M_{\rm Sgr A^*}}{M} \right)^{1/6} \, \text{,}
    \label{eq:vej}
\end{equation}
where \textcolor{black}{$M= m_{\rm s} + m_{\rm p}=(1+q)m_{\rm p}$ is the total mass of the progenitor binary, $m_{\rm c}= M - m_{\rm ej}$ is the mass of the non-HVS member of the former binary that remains bound to Sgr A*}, and $M_{\rm Sgr A^*}=4\times10^{6} \, \mathrm{M_\odot}$ \citep{Eisenhauer2005, Ghez2008}. We assume that stars are ejected from the GC at a constant rate $\eta$ and that the mass of Sgr A* remains unchanged with time. We choose $\eta=10^{-4} \, \mathrm{yr^{-1}}$ \citep[see][]{Brown2015rev} as our fiducial ejection rate.

Our present-day mock ejected star population consist of stars ejected $t_{\rm ej}$ ago that are not yet stellar remnants. We assume that the GC has been ejecting stars without pause since its time of formation, taken here to be shortly after the Big Bang approximately $13.8 \, \mathrm{Gyr}$ ago \citep{Planck2020}. We assign $t_{\rm ej}$ uniformly: 
\begin{align}
    t_{\rm ej} &= \epsilon_1 \cdot \mathrm{13.8 \, Gyr}  \; ,
\label{eq:tflight}
\end{align}
where $0<\epsilon_1 < 1$ is a uniform random number. In practice, only HVSs ejected less than $\sim50 \, \mathrm{Myr}$ ago will be close enough (and thus bright enough) to be assigned a radial velocity in any current or future \textit{Gaia} data release. We assume both stars in the binary reach ZAMS at the same time. Each star has a maximum lifetime $t_{\rm life}$, taken here as the elapsed time necessary for a star to evolve from \textit{ZAMS to the moment it first becomes a stellar remnant}. \textcolor{black}{We assume that at ejection there is no preference for older or younger HVSs, and therefore we say the age of the binary at ejection $t_{\rm age, ej}$ is a random fraction} $\epsilon_2$ of the maximum total lifetime of the binary;
\begin{equation}
    t_{\rm age, ej} = \epsilon_2 \cdot t_{\rm max} \; ,
\end{equation}
where $t_{\rm max} \equiv \mathrm{min}[t_{\rm life}(m_{\rm c}), t_{\rm life}(m_{\rm ej}), \mathrm{13.8 \, Gyr}]$ to ensure that a) both stars are non-remnants at the time of ejection, and b) the binary is not older than the Universe, as $t_{\rm life}>13.8 \, \mathrm{Gyr}$ is often true of low-mass stars. We calculate $t_{\rm life}$ for each ejected star using the single stellar evolution \texttt{SSE} algorithms of \citet{Hurley2000} as implemented within The Astrophysical MUltipurpose Software Environment, or \texttt{AMUSE}\footnote{\url{https://amuse.readthedocs.io/en/latest/index.html}} \citep{Portegies2009,Portegies2013,Pelupessy2013,Portegies2018}. 

After ejection, the remaining lifetime of the star $t_{\rm left}$ is
\begin{equation}
    t_{\rm left} = t_{\rm life}(m_{\rm ej}) - t_{\rm age, ej} . \; 
\end{equation}
We remove stars for whom $t_{\rm ej}>t_{\rm left}$, i.e. stars which are remnants at the present day. The flight time of each surviving mock ejected star is then
\begin{equation}
    t_{\rm flight} = t_{\rm ej}
\end{equation}
and its current age is
\begin{equation} \label{eq:tage}
    t_{\rm age,0} = t_{\rm age,ej} + t_{\rm flight} \; .
\end{equation}

\subsection{Orbital integration} \label{sec:methods:potential}

We assume that the \textcolor{black}{the Hills mechanism ejects stars isotropically away from Sgr A*. We therefore eject HVSs in random directions, initializing them on random points on the surface of the Sgr A* sphere of influence $3 \, \mathrm{pc}$ in radius \citep{Genzel2010} with initial velocities pointing radially away from the GC}. We then propagate the stars through the Milky Way using the Galactic potentials of \citet{McMillan2017}, who fit a many-component potential to various kinematic data using a Monte Carlo Markov Chain (MCMC) method. For each realization in which we eject \textcolor{black}{stars from the GC, we draw a Solar position and velocity from among the \citet{McMillan2017} MC chain (P. McMillan, private communication) as well as a Galactic potential.} \textcolor{black}{Using a fifth-order Dormand-Prince integrator \citep{Dormand1980} and a timestep of $0.1 \, \mathrm{Myr}$, w}e integrate ejected star trajectories through this potential using the \texttt{PYTHON} package \texttt{GALPY}\footnote{\url{https://github.com/jobovy/galpy}} \citep{Bovy2015}. 

\subsection{Mock photometric observations} \label{sec:methods:observations}
We determine the current luminosity, effective temperature, radius and surface gravity \textcolor{black}{for each ejected star} using the \texttt{SSE} models implemented within \texttt{AMUSE}. We also identify each star's current evolutionary stage (e.g. main sequence, red giant, core helium-burning) adopting the conventions of \citet{Hurley2000} to designate stages. Next, we calculate the visual extinction $A_{\rm V}$ at each star's distance and sky position using the \texttt{MWDUST}\footnote{\url{https://github.com/jobovy/mwdust}} three-dimensional Galactic dust map \citep{Bovy2016} assuming a \citet{Cardelli1989} reddening law with $R_{\rm V}=3.1$. From each star's temperature, surface gravity and visual extinction, we obtain mock photometric observations using the MESA Isochrone and Stellar Tracks, or \texttt{MIST} \citep{Dotter2016, Choi2016} model grids\footnote{\url{https://waps.cfa.harvard.edu/MIST/}}. We interpolate the appropriate bolometric correction tables to determine each star's apparent magnitude in the \textit{Gaia} EDR3 $G$ and $G_{\rm RP}$ bands\footnote{\url{see https://www.cosmos.esa.int/web/gaia/edr3-passbands}} \citep{Riello2021} as well as the Johnson-Cousins $V$ and $I_{\rm c}$ bands \citep{Bessell1990}. The apparent magnitude in the \textit{Gaia G$_{\rm RVS}$} band can then be computed from the $G$, $V$ and $I_{\rm c}$ magnitudes using the polynomial fits in \citet{Jordi2010} (table 3). To select stars which would appear in the S$^5$ survey, we determine each star's apparent magnitude in the
Dark Energy Camera (DECam) $g$ and $r$ filters \citep{DES2018} as well.

\subsection{Identifying \textit{Gaia}-visible HVSs}

With apparent magnitudes computed, we next identify which stars would appear as promising HVS candidates in various \textit{Gaia} data releases. As in \citetalias{Evans2022}, these stars must satisfy three criteria:

\begin{itemize}
    \item They must satisfy the apparent magnitude and effective temperature conditions (described below) to appear in the radial velocity catalogue of a given data release.
    \item Their mock relative parallax uncertainties must be <20\%, otherwise distance estimation becomes non-trivial \citep[see][]{BailerJones2015}.
    \item When comparing its Galactocentric velocity to the Galactic escape velocity at its position according to the best-fitting potential of \citet{McMillan2017}, it must have an >80\% chance of being unbound to the Galaxy when sampling over its astrometric and radial velocity uncertainties. 

\end{itemize}    

These above cuts match closely those used to search for HVS candidates in  \textit{Gaia} DR2 \citep{Marchetti2019} and EDR3 \citep{Marchetti2021}. For concision, when we use the terms `\textit{Gaia} DR2/(E)DR3/DR4' we are referring exclusively to the \textcolor{black}{subsets of these data releases with measured radial velocities}, and by the term `HVS' we refer only to those stars which satisfy these criteria.

The faint-end magnitude limit for the \textit{Gaia} DR2 radial velocity catalogue is $G\simeq12$, though the precise faint-end limit varies on the sky due to the \textcolor{black}{scanning pattern of the \textit{Gaia} satellite itself and due to stellar crowding in source-dense regions such as the GC and Large Magellanic Cloud} \citep[see][]{Boubert2020cogi, Boubert2020cogii}. To more realistically account for these observational realities, we use the \textit{Gaia} DR2 spectroscopic selection function as estimated by \citet{Everall2022}\footnote{see \url{https://github.com/gaiaverse/selectionfunctions}}. \textcolor{black}{The selection function assigns each star has a probability $p$ of appearing in the \textit{Gaia} DR2 radial velocity catalogue depending on its sky position, brightness and colour. We classify an HVS as `DR2-detectable' if a uniform random number} $0<\epsilon<1$ satisfies $\epsilon<p$. \textcolor{black}{The \textit{Gaia} DR2 spectroscopic pipeline from providing validated radial velocities for only for sources with effective temperature ranges in the range $3500 \, \mathrm{K} \leq  T_{\rm eff} \leq 6900 \, \mathrm{K}$ \citep{Katz2019}. Mock ejected stars with effective temperatures outside this range are removed from our \textit{Gaia} DR2-detectable sample.} 

\textcolor{black}{We estimate the \textit{Gaia} DR2 astrometric uncertainties for each star using the DR2 astrometric spread function of \citet{Everall2021cogiv}\footnote{see \url{https://github.com/gaiaverse/scanninglaw}}. The astrometric spread function computes the full 5-D covariance matrix for each source, providing uncertainties and correlations among the position, parallax and proper motion.} We estimate radial velocity errors for each star \textcolor{black}{based on its $V$-band magnitude and spectral type} using the \texttt{PYTHON} package \texttt{PyGaia}\footnote{\url{https://github.com/agabrown/PyGaia}}, and assume for all stars that the radial velocity uncertainties are uncorrelated to the astrometric uncertainties.

We follow a similar procedure to identify high-confidence mock HVS candidates appearing in \textit{Gaia} Early Data Release 3. \textit{Gaia} EDR3 parallax uncertainties are improved by 30 per cent relative to DR2, and proper motion uncertainties improve by a factor of 2 \citep{Gaia2020EDR3}. \textcolor{black}{EDR3 however, does not provide new or updated radial velocities. \textit{Gaia} DR2 radial velocity measurements have been simply ported to their EDR3 counterparts.} We once again use the DR2 spectroscopic selection function of \citet{Everall2022} to select stars appearing in this catalogue, the DR3 astrometric spread function of \citet{Everall2021cogiv} to assign astrometric uncertainties and \texttt{PyGaia} to assign radial velocity uncertainties.

\textcolor{black}{The full \textit{Gaia} DR3, released 13 June 2022, contains the EDR3 astrometric solutions as well as radial velocity measurements}\footnote{\textcolor{black}{This work was initially submitted for publication prior to the release of \textit{Gaia} DR3. In \citet{Marchetti2022} we search for HVS candidates within DR3 and use the methodology of this work to infer updated constraints on the GC environment.}} for $\sim$33 million sources brighter than $G_{\rm RVS}\simeq14$ in the effective temperature range $[3500\, \mathrm{K}, 6900 \, \mathrm{K}]$ \citep{Katz2019}. Improvements in the \textit{Gaia} spectroscopic pipeline\footnote{see \url{https://www.cosmos.esa.int/web/gaia/dr3}} additionally allow validated radial velocity measurements for $7000 \, \mathrm{K} < T_{\rm eff} < 14500 \, \mathrm{K}$ sources to a depth of $G_{\rm RVS}\lesssim12$. \textcolor{black}{We use these same criteria to select stars detectable in DR3, since more detailed information about the DR3 spectroscopic selection function is not yet available.} The DR3 astrometric spread function of \citet{Everall2022} and \texttt{PyGaia} are once again used to assign astrometric and radial velocity uncertainties, respectively.

Finally, we identify stars visible in the fourth and (nominally) final \textit{Gaia} data release. Radial velocities will be available for sources cooler than $6900 \, \mathrm{K}$ \textcolor{black}{and brighter than} the $G_{\rm RVS}=16.2$ mag limiting magnitude of the \textit{Gaia} radial velocity spectrometer \citep{Cropper2018, Katz2019}. For hotter stars, \textcolor{black}{We make the assumption that} validated radial velocities will be available for sources brighter than $G_{\rm RVS}=14$. \textcolor{black}{We use the astrometric covariance matrix as computed using the \citet{Everall2021cogiv} astrometric spread function to estimate the DR4 astrometric covariance matrix. We scale down the diagonal elements of the matrix (corresponding to the astrometric errors) according to the predicted \textit{Gaia} performance --} relative to DR3, parallax precisions in DR4 will improve by $\sim$33\% and proper motion precisions by $\sim$80\%\footnote{\url{https://www.cosmos.esa.int/web/gaia/science-performance}, see also \citet{Brown2019}.}. \textcolor{black}{We assume that off-diagonal elements, corresponding to the correlations between the astrometric errors, remain unchanged from their DR3 estimations.}

\begin{figure*}
    \centering
    \includegraphics[width=0.65\columnwidth]{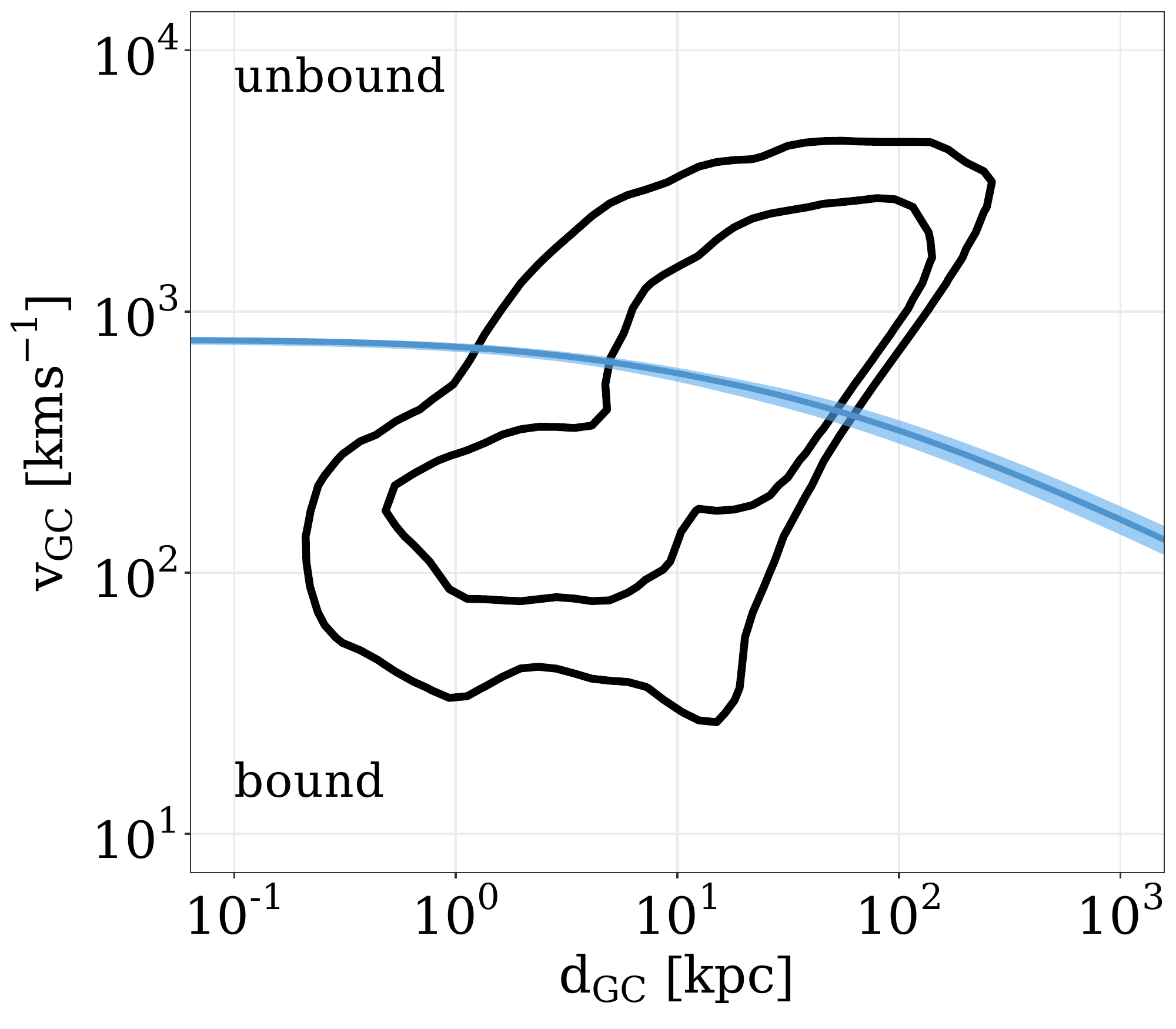}
    \includegraphics[width=0.65\columnwidth]{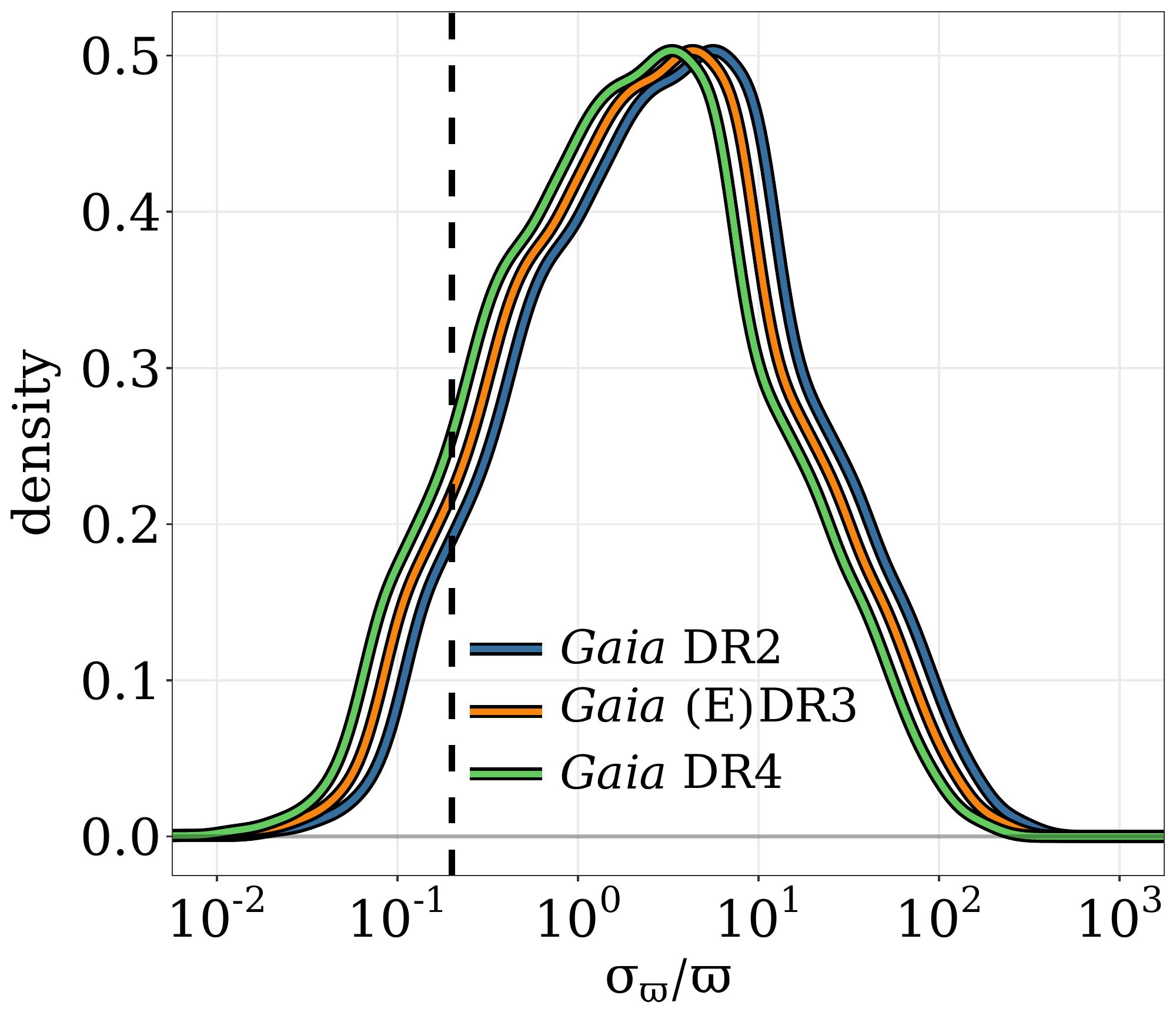}
    \includegraphics[width=0.65\columnwidth]{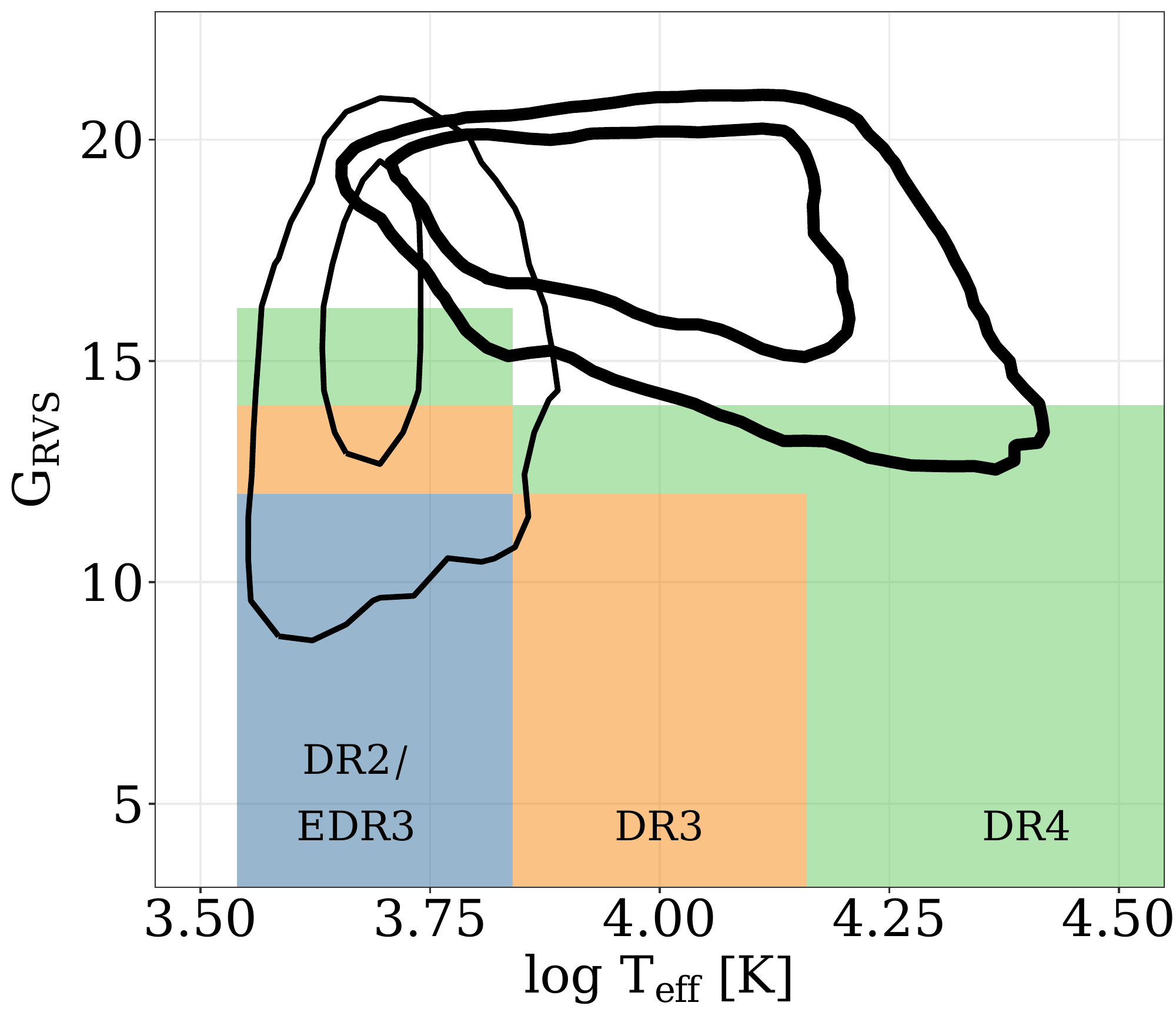}
    \caption{\textit{Left:} 1$\sigma$ and 2$\sigma$ contours of the distribution of Hills mechanism-ejected stars in Galactocentric velocity vs. distance.  The blue curve and shaded region shows the median $\pm$1$\sigma$ spread of the escape velocity to infinity from each Galactocentric distance in the \citet{McMillan2017} family of Galactic potentials. \textit{Centre:} The distribution of the relative \textit{Gaia} parallax uncertainty $\sigma_\varpi / \varpi$ for mock ejected stars. The blue curve shows \textit{Gaia} DR2 uncertainties and the orange and green curves show predicted (E)DR3 and DR4 performance, respectively. The vertical dashed line shows $\sigma_\varpi / \varpi = 0.2$. \textit{Right:} Stars must lie in the shaded regions of $G_{\rm RVS}$ vs. $T_{\rm eff}$ to be assigned a validated radial velocity in \textit{Gaia} DR2, (E)DR3 and DR4. The thick lines show the 1$\sigma$ and 2$\sigma$ density contours for the distribution of main sequence ejected stars in this space. The thin lines show the distribution of post-main sequence stars. In all three panels, only $t_{\rm flight}\leq 100 \, \mathrm{Myr}$ stars brighter than $G=20.7$ stars are shown.}
    \label{fig:GRVSepar}
\end{figure*}

In Fig. \ref{fig:GRVSepar} we illustrate how restrictive our cuts are. After propagation, the left panel shows 68\% and 95\% density contours of Galactocentric distances and velocities for stars ejected from the GC in our fiducial model. The escape velocity curve from the best-fit potential of \citet{McMillan2017} and 1$\sigma$ scatter is shown with the blue line. To keep focus only on stars with a non-negligible chance of being detected by \textit{Gaia}, we show only stars ejected less than 100 Myr ago that are bright enough to appear in the \textit{Gaia} source catalogue, i.e. brighter than the 20.7th magnitude in the \textit{Gaia G} band \citep{Gaia2016Prusti}. Among stars ejected from the GC, there are two distinct populations: a population of stars which remain bound within the inner $\sim10 \, \mathrm{kpc}$ of the Galaxy with velocities of tens to hundreds of $\mathrm{km \ s^{-1}}$, and a population unbound to the Galaxy extending to large distances with velocities of $\gtrsim1000 \, \mathrm{km \ s^{-1}}$.  Less than half of ejected stars are unbound to the Galaxy. While valuable information about the Galactic potential and the GC stellar environment is also encoded in stars that are ejected at large-but-not-unbound velocities, unbound stars are easier to identify as promising HVS candidates\footnote{A proviso: \citet{Brown2007} had success finding \textit{bound} HVS candidates by searching for early-type stars at high Galactic latitudes and large heliocentric distances $d\gtrsim10 \, \mathrm{kpc}$.} and their origins are less ambiguous -- they are the focus of this paper.  In the centre panel we show how the relative \textit{Gaia} parallax error $\sigma_\varpi / \varpi$ for our mock HVS populations improves with each \textit{Gaia} data release. The dashed vertical line shows our parallax error cut at $\sigma_\varpi / \varpi=0.2$. Only $6$\%, $8$\% and $11$\% of HVSs will satisfy $\sigma_\varpi / \varpi<0.2$ in \textit{Gaia} DR2, (E)DR3 and DR4 respectively. The majority of ejected HVSs will have a relative parallax error $\gg1$. In the right panel, the shaded regions show the effective temperature and (approximate) $G_{\rm RVS}$ limits of the \textit{Gaia} DR2/EDR3, DR3 and DR4 radial velocity catalogues. In this space, we show the 68\% and 95\% density contours for the distributions of main sequence (thick lines) and post-main sequence (thin lines) ejected star populations separately. Overall, main sequence HVSs in the \textit{Gaia} source catalogue are too dim and too hot to appear in \textit{Gaia} DR2/EDR3 and DR3, but an appreciable population may be found in DR4 (see also \citetalias{Evans2022}). Evolved HVSs fare much better, however. Although they constitute only $\sim 8$\% of total ejected HVSs, a large fraction are sufficiently cool and bright to be assigned radial velocities in \textit{Gaia} DR2/(E)DR3/DR4 in principle. 

\section{Results} \label{sec:results}

\begin{figure*}
    \centering
    \includegraphics[width=1.8\columnwidth]{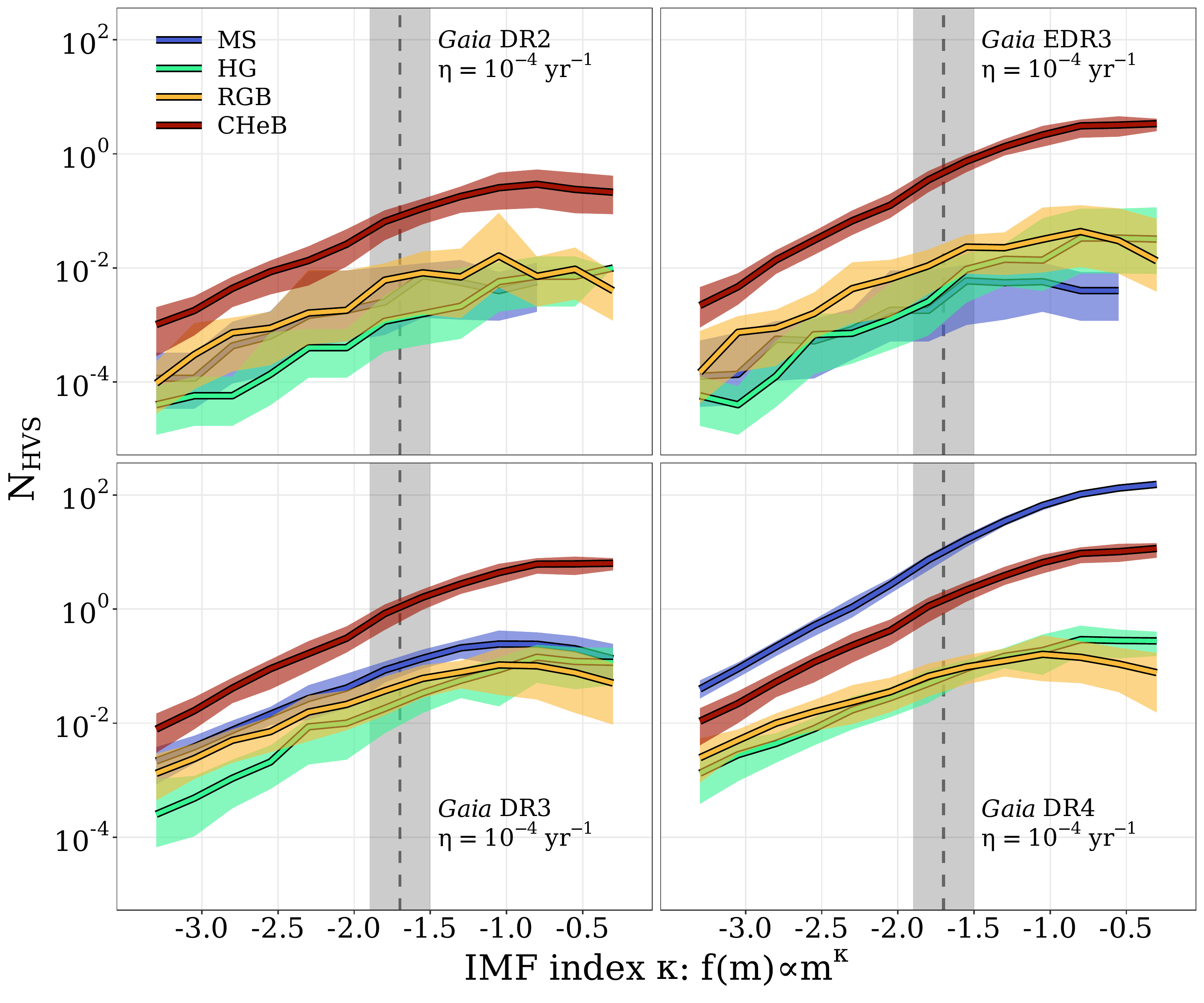}
    \caption{The number $N_{\rm HVS}$ of high-confidence detectable HVSs in the radial velocity catalogues of \textit{Gaia} DR2, EDR3, DR3 and DR4 (clockwise from top-left) plotted against the HVS progenitor binary IMF index $\kappa$ for a fixed HVS ejection rate of $10^{-4} \, \mathrm{yr^{-1}}$. HVS populations are split into main sequence (MS; blue), Hertzsprung gap/subgiant branch (HG; green), red giant branch (RGB; gold) and core helium-burning (CHeB; red) populations. Shaded regions span the 16th to 84th percentiles over 5000 iterations. The vertical dashed line and shaded region show $\kappa=-1.7\pm0.2$ \citepalias{Lu2013}.}
    \label{fig:NHVSkappa}
\end{figure*}

\subsection{The evolved HVS population}

Having explored solely the main sequence HVS population in \citetalias{Evans2022}, here we first describe the number of high-confidence HVSs of all evolutionary stages we predict to appear in current and future \textit{Gaia} data releases. 

We showed in \citetalias{Evans2022} that the number of HVSs appearing in \textit{Gaia} depends most critically on the assumed IMF slope $\kappa$ and the HVS ejection rate $\eta$. In Fig. \ref{fig:NHVSkappa} we show how the number of HVSs depends on $\kappa$ in each of \textit{Gaia} DR2, EDR3, DR3 and DR4. We split HVSs into main sequence (MS), subgiant branch or Hertzsprung gap (HG), red giant branch (RGB) and core helium-burning (CHeB) phases. The dotted vertical line shows our fiducial assumption for $\kappa$, and the shaded region shows a $\pm0.2$ uncertainty applied \citepalias{Lu2013}. These are estimates for our fiducial HVS ejection rate of $10^{-4} \, \mathrm{yr^{-1}}$ -- they can be scaled linearly up or down for other ejection rates \citepalias[c.f.][fig. 2]{Evans2022} since we assume a constant ejection rate. At this ejection rate, less than one HVS in total is expected in \textit{Gaia} DR2 unless the IMF \textcolor{black}{of HVS progenitor binaries} is very top-heavy. Core helium-burning stars are the most likely to appear in this data release, while main sequence, Hertzsprung gap and red giant branch HVSs are all quite rare and more or less equally likely. For \textit{Gaia} EDR3, $\geq1$ evolved HVSs are expected to appear in this survey if the GC IMF is more top-heavy than our fiducial assumption. Given that no HVS has yet been detected in EDR3, this fact can already place some meaningful constraints on the GC stellar environment. Once again, if HVSs \textit{were} likely to lurk in this data release, they would most likely be core helium-burning.

\textcolor{black}{Looking ahead to subsequent releases, ${1.1_{-0.9}^{+1.5}}$ HVS are} expected in \textit{Gaia} DR3 given our fiducial assumptions, most likely a core helium-burning HVS. If the IMF of HVS progenitors were to be particularly top-heavy, several core-helium burning would be expected and the probability of detected a main sequence HVS becomes non-insignificant.

Finally for DR4, we predict $10.9_{-4.2}^{+4.8}$ HVSs for our fiducial model. Unlike earlier data releases, the main sequence HVS population will outnumber evolved ones. Of the $1.4_{-0.7}^{+1.6}$ evolved stars we expect to appear in this data release in our fiducial model, $1.3_{-0.8}^{+1.7}$ will be horizontal branch stars with typical masses of $\sim5 \, \mathrm{M_\odot}$. Unless the IMF is quite top-heavy or the HVS ejection rate quite high, we expect fewer than one hypervelocity Hertzsprung gap or red giant star to appear in DR4. It may seem counter-intuitive that predictions for Hertzsprung gap and red giant branch stars are similar throughout Fig. \ref{fig:NHVSkappa} when the Hertzpsrung gap is known to be a short phase of evolution. It is helpful to note, however, that 96\% of detectable HVSs are more massive than $\sim2 \, \mathrm{M_\odot}$, for whom the red giant branch phase is quite short since their helium core is non-degenerate when it reaches the base of the red giant branch.

\begin{figure*}
    \centering
    \includegraphics[width=0.85\columnwidth]{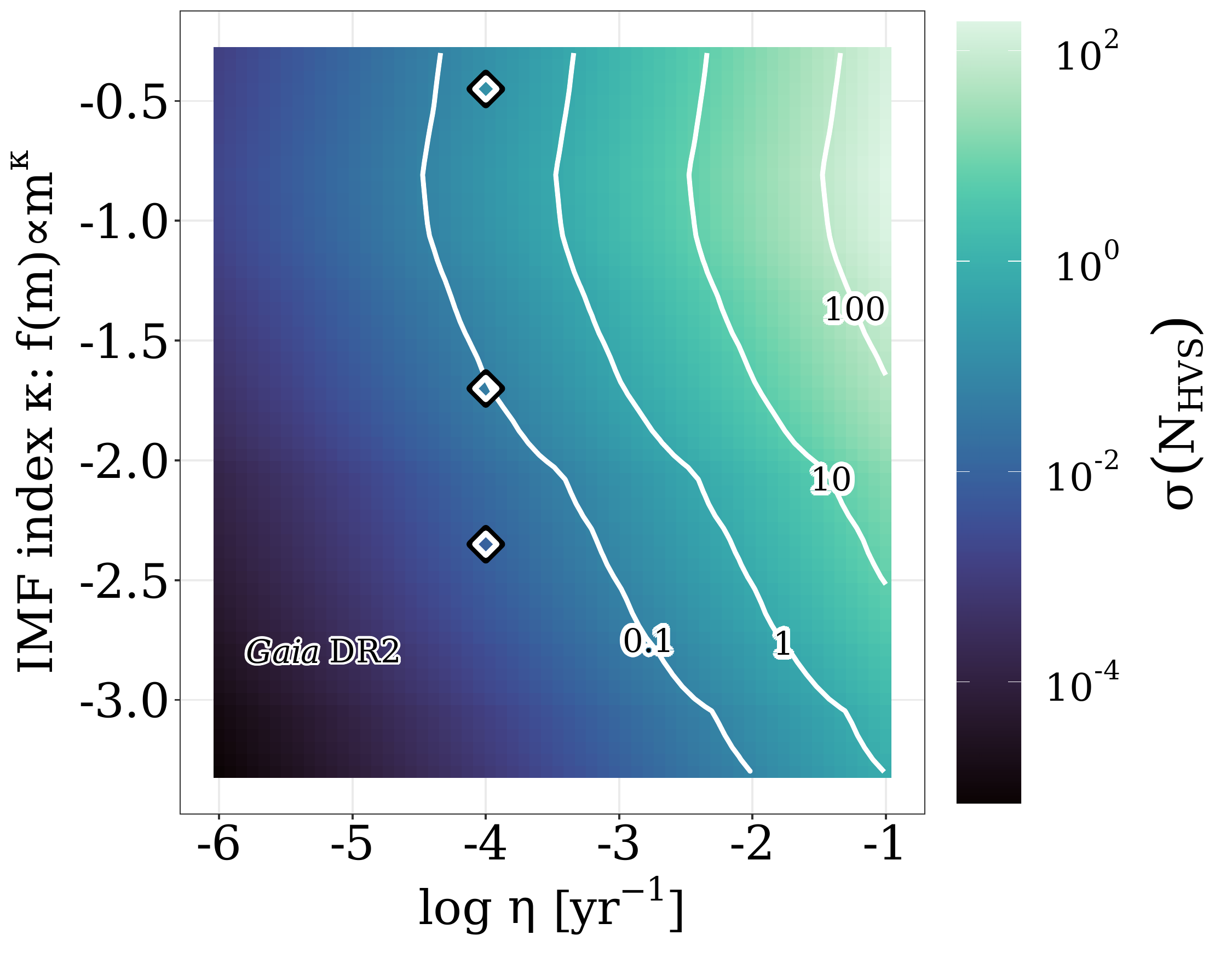}
    \includegraphics[width=0.85\columnwidth]{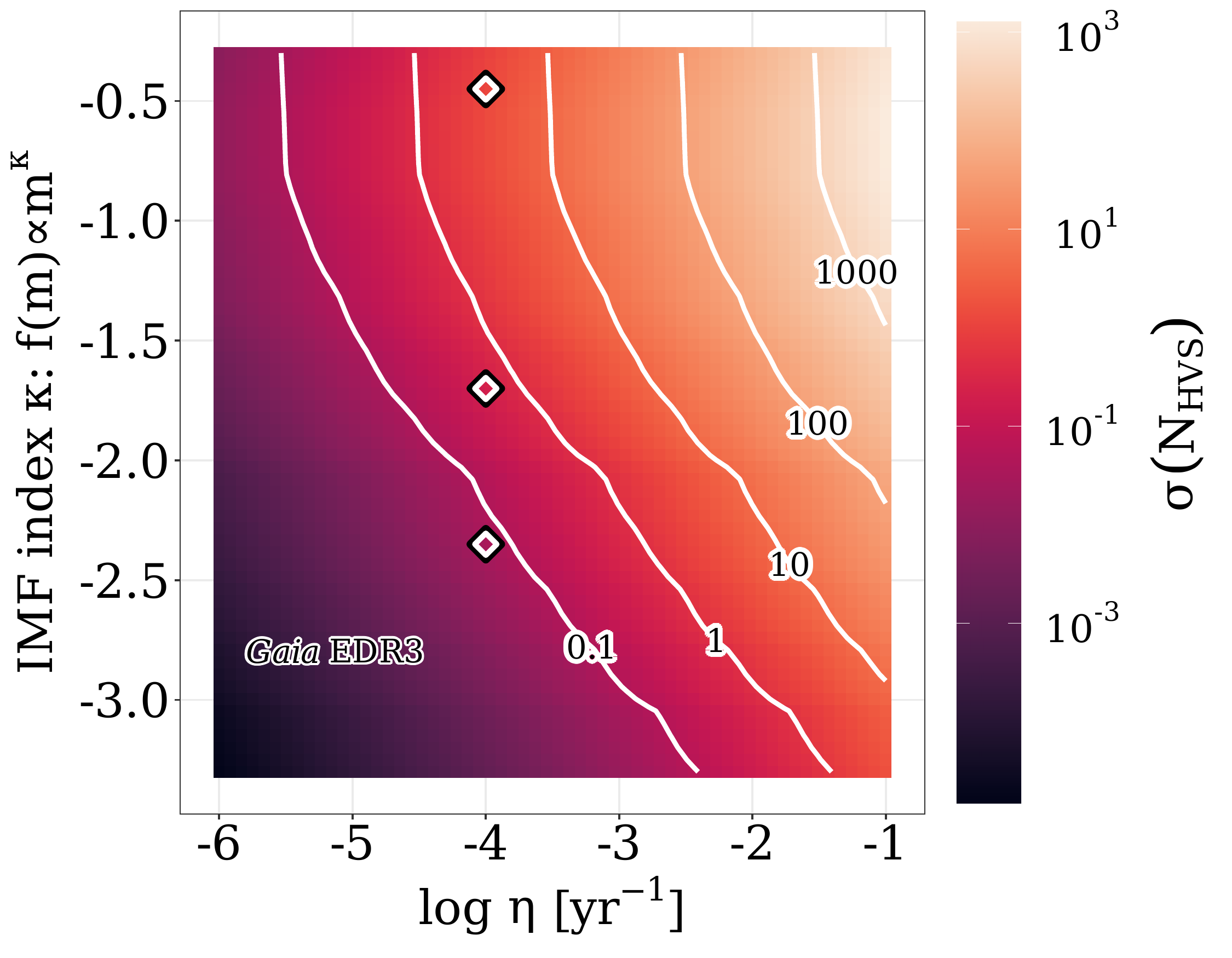}
    \includegraphics[width=0.85\columnwidth]{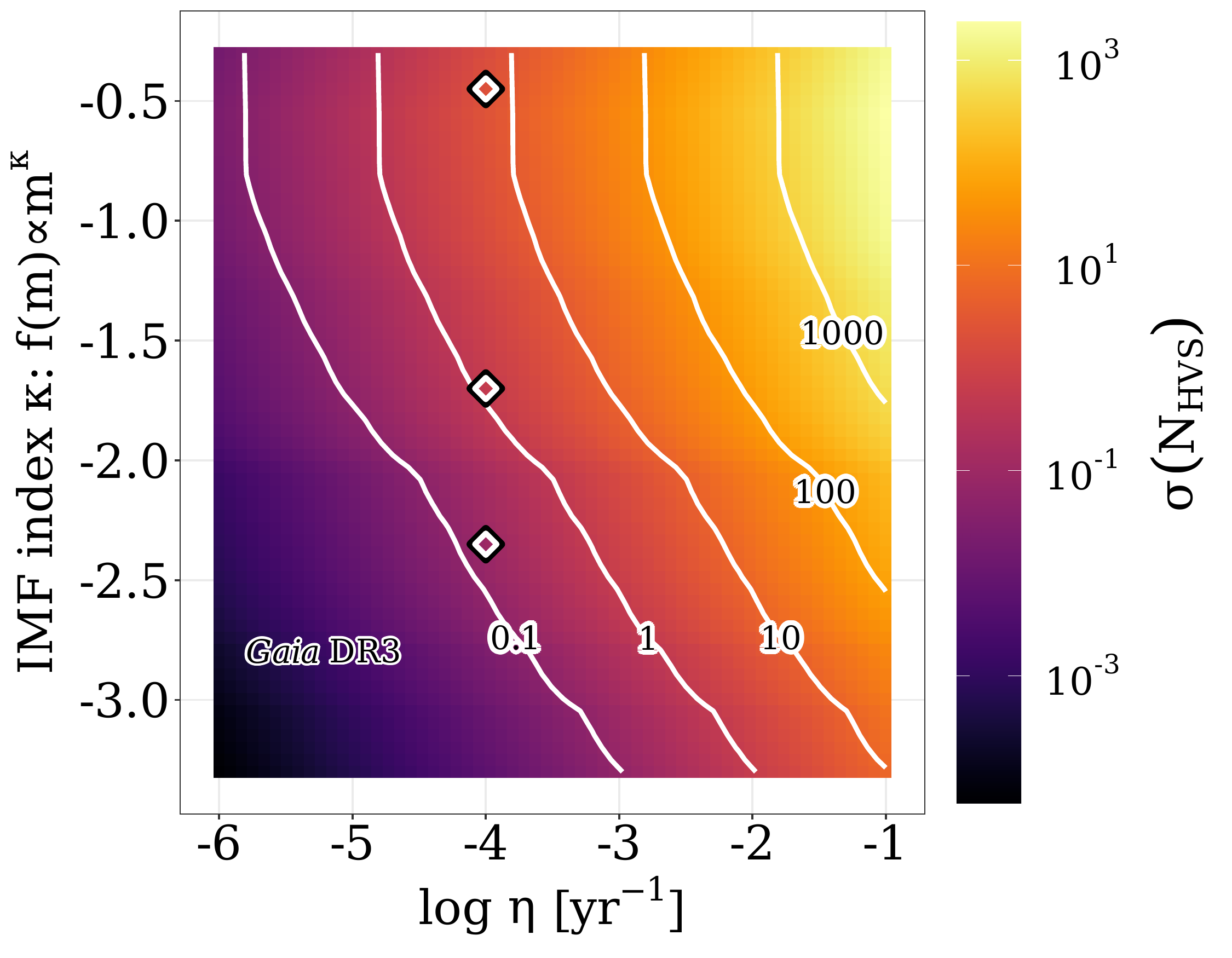}
    \includegraphics[width=0.85\columnwidth]{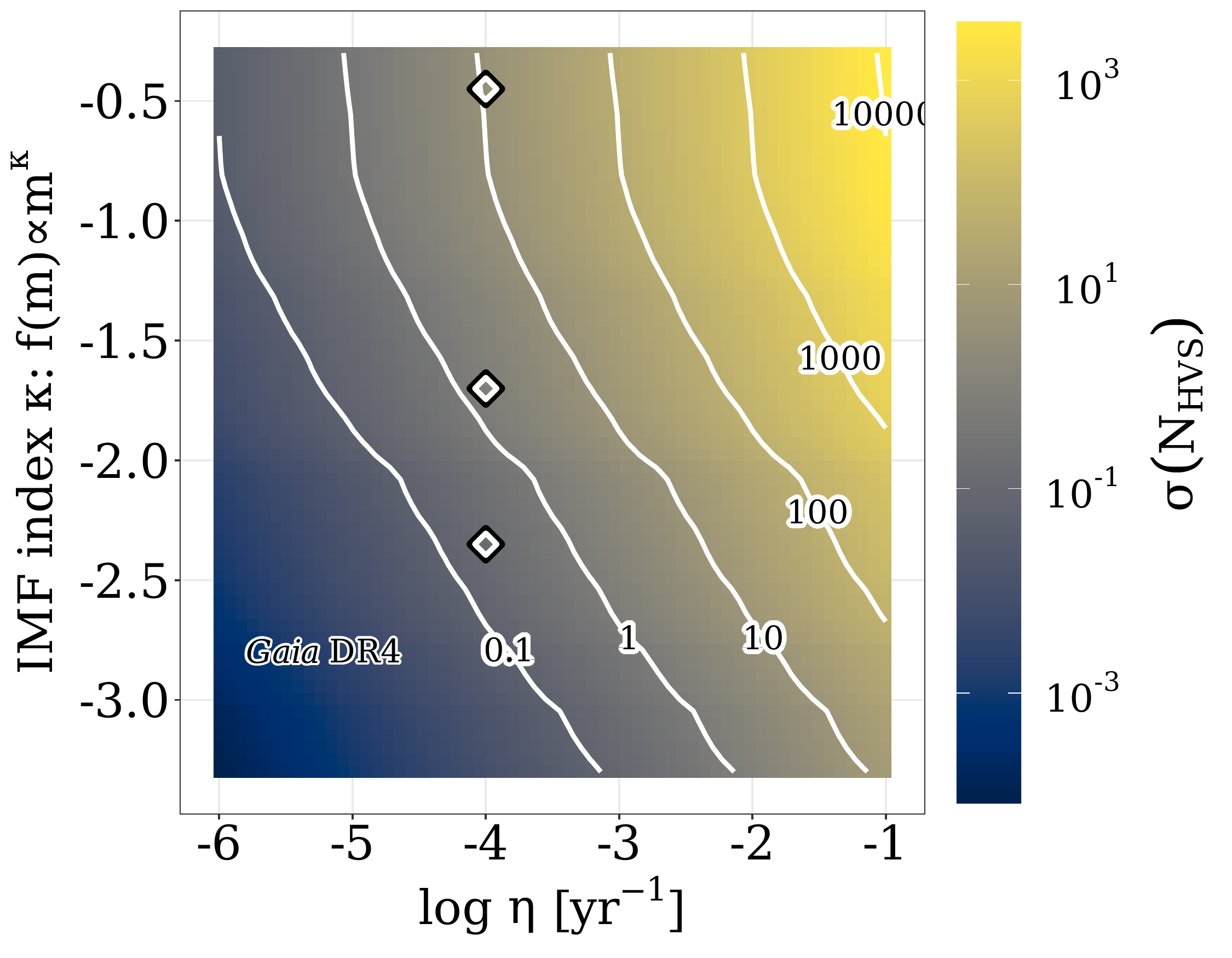}

    \caption{Contour lines show how the the numbers $N_{\rm HVS, \, evolved}$ of evolved HVSs in \textit{Gaia} DR2 (top left), EDR3 (top right), DR3 (bottom left) and DR4 (bottom right) change in the 2D parameter space of the IMF power law index $\kappa$ and the HVS ejection rate $\eta$, averaged over 5000 realizations and smoothed over the grid. The colourbar shows how the 1$\sigma$ scatter of $N_{\rm HVS, \, evolved}$ changes in this space. The black-and-white diamonds indicate our fiducial ejection rate of $\eta=10^{-4} \, \mathrm{yr^{-1}}$ and fiducial $\kappa=-1.7$ \citepalias{Lu2013}, as well as $\kappa=-2.35$ \citepalias{Salpeter1955} and $\kappa=-0.45$ \citepalias{Bartko2010}.}
    \label{fig:HVSkappaeta}
\end{figure*}

When the HVS ejection rate is left as a free parameter, we show with white contour lines in Fig. \ref{fig:HVSkappaeta} how of the number $N_{\rm HVS, \, evolved}$ of post-main sequence HVSs changes in the $\kappa-\eta$ space for each different \textit{Gaia} data release. The colourscales indicate the 1$\sigma$ scatter of $N_{\rm HVS, \, evolved}$ over 5000 iterations smoothed over the grid. The black-and-white diamonds indicate our fiducial $\kappa=-1.7$ \citepalias{Lu2013} and $\eta=10^{-4} \, \mathrm{yr^{-1}}$, as well as models with $\kappa$=-2.35 \citepalias{Salpeter1955} and $\kappa$=-0.45 \citepalias{Bartko2010} for comparison. Similar plots for the main sequence HVS population can be found in \citetalias{Evans2022}, c.f. fig. 4. There are regions of $\kappa-\eta$ space that predict at least $\sim$tens or $\sim$hundreds of high-confidence evolved HVS in \textit{Gaia} DR2 and EDR3. This, however, would be in contradiction to the apparent complete absence of these objects in these data releases. For particularly top-heavy IMFs and reasonably high ejection rates, $\sim$hundreds of evolved HVSs could be found. DR4 will only grow the evolved HVS population by a modest amount. This is in stark contrast to the main sequence population, which can increase in size by more than two orders of magnitude from DR3 to DR4. In Sec. \ref{sec:discussion:curios} we discuss interesting hypervelocity subpopulations in greater detail.

\begin{figure*}
    \centering
    \includegraphics[width=1.8\columnwidth]{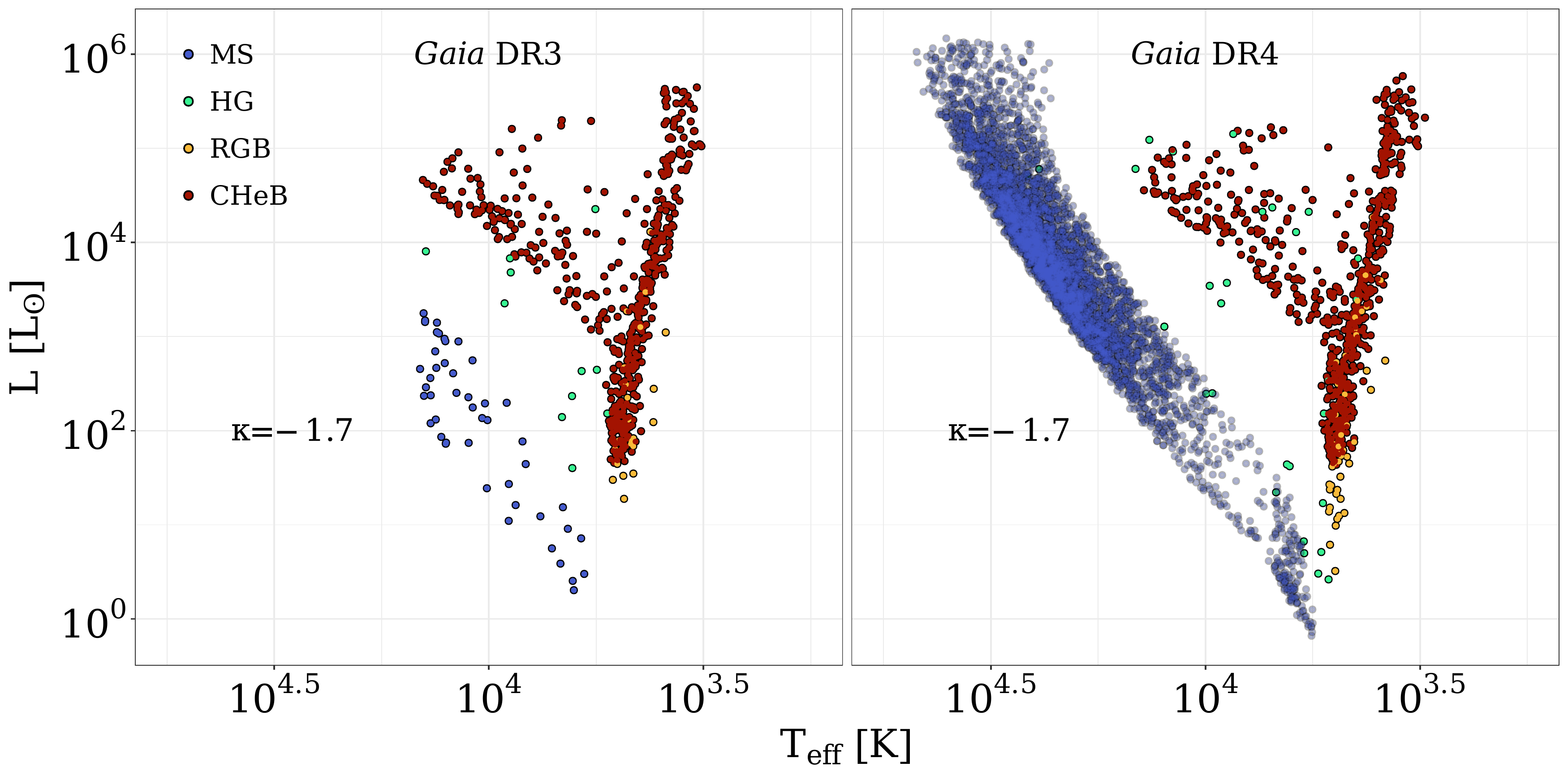}
    \caption{Hertzsprung-Russell diagram for high-confidence detectable HVSs in the radial velocity catalogues of \textit{Gaia} DR3 (left) and DR4 (right) for a fixed IMF slope ($\kappa=-1.7$). Populations are coloured by evolutionary stage as in Fig. \ref{fig:HVSnumbers}. Results are shown stacked over 500 iterations. Main sequence HVSs in DR4 are plotted with semitransparency to better show the distribution.}
    \label{fig:HR}
\end{figure*}

Given reasonable (if optimistic) assumptions, the \textit{Gaia} DR3 and DR4 HVS populations could conceivably be large enough to obtain useful summary statistics. To examine these potential populations in greater detail, in Fig. \ref{fig:HR} we show how the populate the Hertzsprung-Russell diagram stacked over 5000 iterations with our fiducial $\kappa$. In DR3, main sequence HVSs up to $\sim6 \, \mathrm{M_\odot}$ will in principle be present. Detectable core helium-burning stars in this data release would most likely be $m\simeq2-3 \, \mathrm{M_\odot}$ stars in the red clump as well as core helium-burning stars during the entirety of a blue loop phase.  These same post-main sequence populations will be detectable in DR4, as well as a sizeable main sequence HVS population with a typical mass of $\sim 8 \, \mathrm{M_\odot}$.

\subsection{Constraining the GC IMF and ejection rate} \label{sec:bayes}

In \citetalias{Evans2022} we used the absence of main sequence HVSs in \textit{Gaia} DR2 and EDR3 to place constraints on the IMF in the GC and the HVS ejection rate. We showed that unless the IMF \textcolor{black}{among the primaries of HVS progenitor binaries} is extremely top-light ($\kappa\lesssim-3$), ejection rates in excess of $3\times10^{-2} \, \mathrm{yr^{-1}}$ are excluded at 1$\sigma$ confidence. With our analysis of evolved HVSs here, we can improve these constraints considerably. In addition, we compute constraints in a more sophisticated way using a Bayesian inference approach. This allows us to use prior information on $\kappa$ and $\eta$ and other observational evidence about Milky Way HVSs to strengthen constraints. We compute the posterior probabilities

\begin{equation}
    p(\theta | D) = \frac{\mathcal{L}(D | \theta) p(\theta)}{p(D)} \; ,
    \label{eq:bayes}
\end{equation}
where $D$ is the observed HVS data, $\theta \equiv (\kappa,\eta)$ are model parameters for the IMF index and HVS ejection rate in the GC environment, $\mathcal{L}(D | \theta)$ is the likelihood of observing the data given the particular model, $p(\theta)$ accounts for prior knowledge on the model parameters, and $p(D)$ is a normalizing constant. The combination of these yields the posterior probability for the model parameters $p(\theta | D)$.

In the subsections below we outline the existing data we consider, the priors we consider, and the resulting posterior constraints on $\kappa$ and $\eta$.

\subsubsection{the data}

A key observation we consider is the lack of unbound HVS candidates with precise astrometry in the radial velocity catalogues of \textit{Gaia} DR2 \citep{Marchetti2019} and EDR3 \citep{Marchetti2021}, and our selection criteria outlined in Sec. \ref{sec:methods} mirrors quality cuts used in these works. While this places competitive constraints on the GC stellar environment by itself, we can improve constraints further by considering the HVS candidate S5-HVS1 \citep{Koposov2020}, a $\sim2.35 \, M_\odot$ star with an apparent magnitude of $16.0$ in the \textit{Gaia G} band, a breakneck Galactocentric velocity of $v_{\rm ej}\simeq \, 1750 \, \mathrm{km \ s^{-1}}$ and a relatively short flight time from the GC of $t_{\rm flight}\simeq 4.8 \, \mathrm{Myr}$. It was identified in a subsurvey of the S$^5$ survey \citep{Li2019}, which as of June 2019 has covered $\sim345$ square degrees with 115 fields observed with the Anglo-Australian Telescope (AAT). It can be stated with confidence that S5-HVS1 is the only HVS within the S$^5$ catalogue -- by identifying S5-HVS1 analogues from our mock populations of HVSs, we can determine which models are consistent with S5-HVS1. To identify S5-HVS1 analogues, we roughly reproduce the S$^5$ selection criteria by taking HVSs within the S$^5$ footprint \citep[see][table 2]{Li2019} which have mock \textit{Gaia} parallaxes satisfying $\varpi < 3\sigma_{\varpi} + 0.2$, mock DECam photometry satisfying $15<g<19.5$ and $-0.4 < (g-r) < +0.1$. Since only stars with radial velocities larger than $800 \, \mathrm{km \ s^{-1}}$ were selected for further inspection \citep{Koposov2020}, we apply this criterion as well.    

\subsubsection{the priors}

We consider two sets of priors on the IMF index in the GC $\kappa$ and the HVS ejection rate $\eta$; one set in which we assume uniform priors across the $\kappa-\eta$ range we explore, and one more restrictive set which considers modern determinations of these parameters. 

For the set of restrictive priors, we assume $\kappa$ is normally distributed with a mean at $\kappa=-1.7$ and a standard deviation of 0.2, following from \citetalias{Lu2013} who simultaneously fit several properties of the young stellar cluster in the inner $0.5 \, \mathrm{pc}$ of the Galaxy. They compare the \textit{Keck K'}-band luminosity function of young stars in the GC \citep{Do2013} to mock observations of synthetic star clusters to determine this IMF slope via a Bayesian inference approach. While not quite as top-heavy as other IMF determinations near the centre of the Galaxy \citep[e.g.][]{Bartko2010}, this is but another indication that the initial mass function, at least among young stellar structures in the GC, is at least modestly top-heavy \citepalias[see][Sec. 3 and references therein]{Evans2022}.

A recent, robust determination of the HVS ejection rate $\eta$ with associated uncertainties does not yet exist. However, reasonable estimates from theoretical modelling \citep{Hills1988, Yu2003}, detailed simulations \citep{Zhang2013}, and calibration to known HVS candidates \citep{Bromley2012, Brown2014, Marchetti2018} and to rates of tidal disruption events \citep[see][]{Bromley2012, Brown2015, Stone2020} support an ejection rate in the range $10^{-5} \, \mathrm{yr^{-1}} - 10^{-3} \, \mathrm{yr^{-1}}$. For our set of restrictive priors we therefore assume a prior of the form

\begin{align}
    p(\mathrm{log} \eta) \propto \, &\mathrm{tanh}[\ell(\mathrm{log} \eta - (-5) - \pi/\ell)] \\ \nonumber
    &- \mathrm{tanh}[\ell(\mathrm{log} \eta - (-3) + \pi/\ell)] \, \mathrm{,}
\end{align}
where $\ell=8$ is a smoothing parameter, such that the prior probability is uniform between $10^{-5}-10^{-3} \, \mathrm{yr^{-1}}$ and quickly and smoothly drops to zero outside this range. We assume the priors on $\kappa$ and $\eta$ are entirely uncorrelated, i.e. $p(\kappa, \eta) = p(\kappa)p(\eta)$. 

\subsubsection{the likelihood}

The likelihoods $\mathcal{L}(D | \theta)$ are computed with an MC approach. For each $(\kappa, \eta)$ combination in our model grid, $\mathcal{L}(D | \theta)$ is the probability computed over 5000 repeated MC realizations that the model simultaneously satisfies zero HVSs being found in \textit{Gaia} EDR3 and the existence one (and only one) S5-HVS1 analogue, where S5-HVS1 analogues are selected as described above. These outcomes are not independent -- if a particular realization results in many EDR3-detectable HVSs, it is likely to produce many S5-HVS1 analogues as well.

\subsubsection{the posteriors}

\begin{figure*}
    \centering
    \includegraphics[width =\columnwidth]{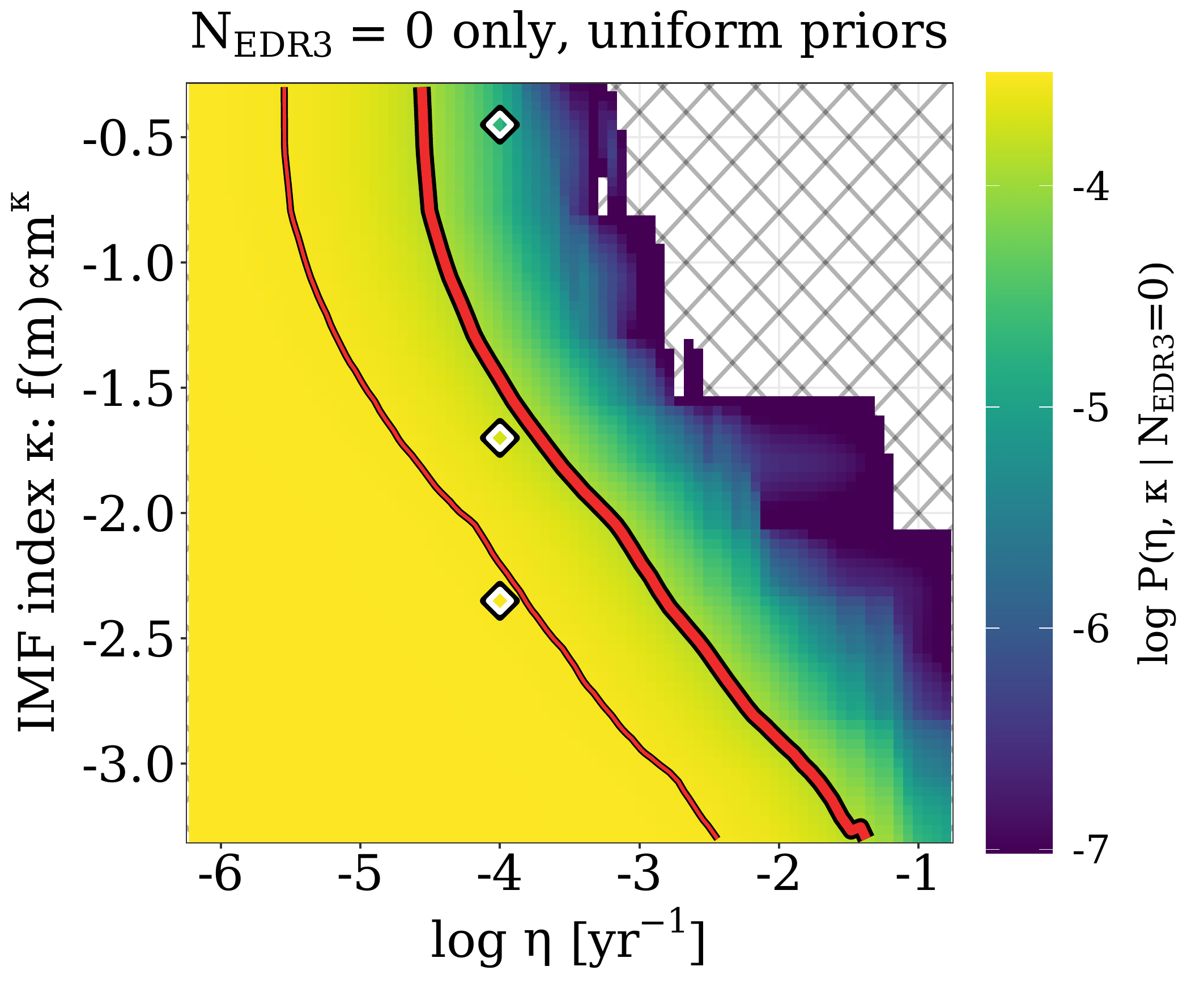}
    \includegraphics[width =\columnwidth]{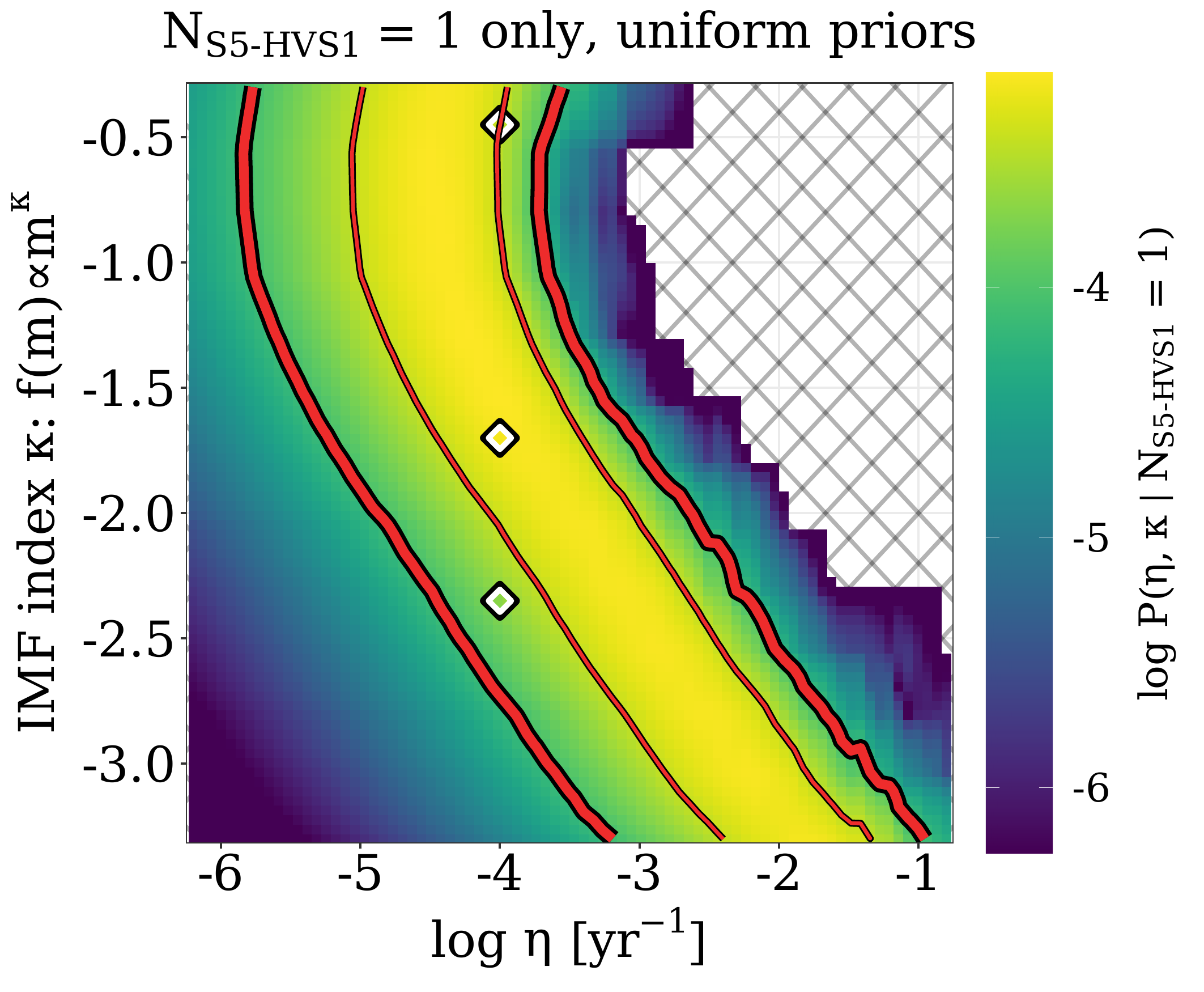}
    \includegraphics[width =\columnwidth]{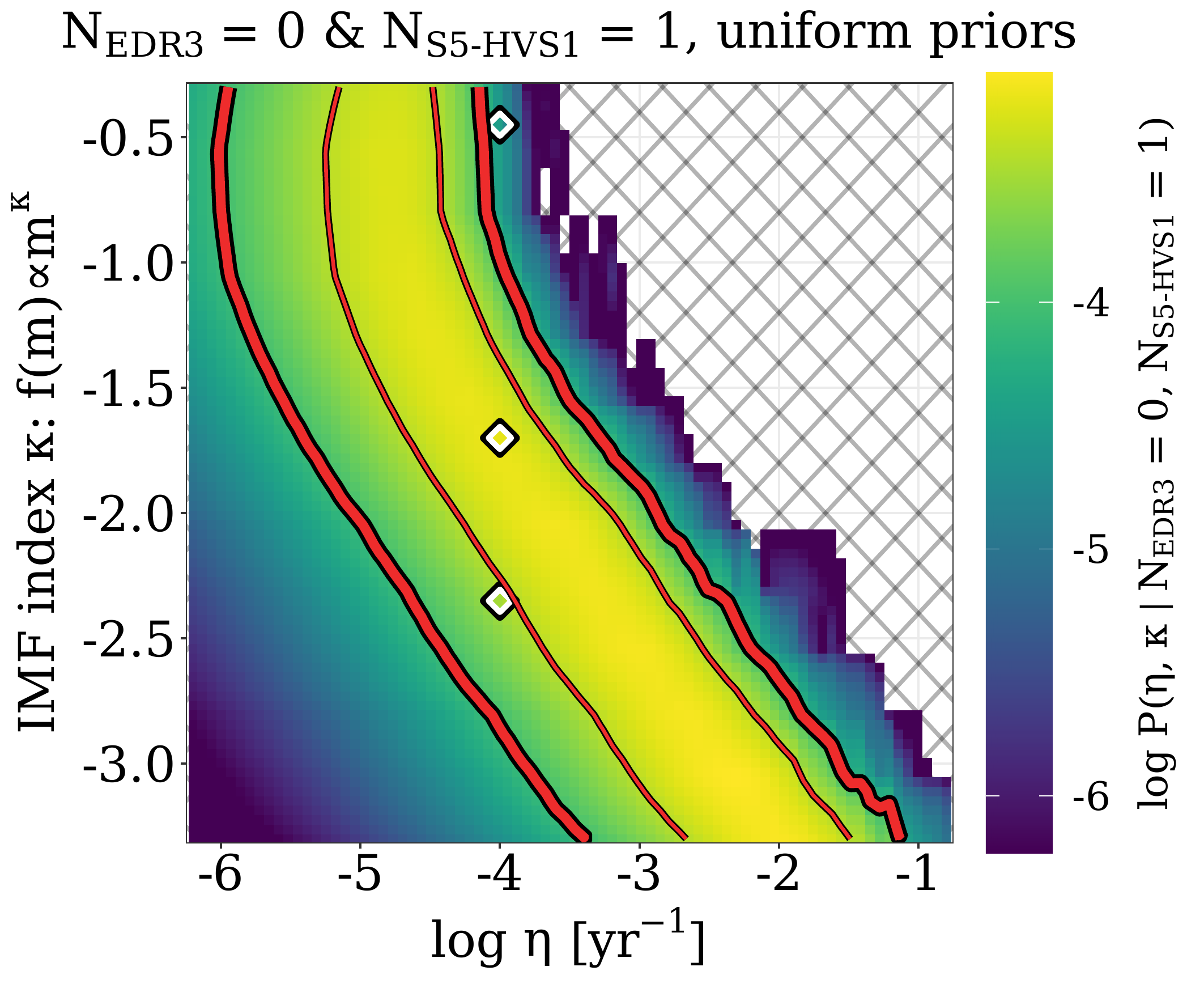}
    \includegraphics[width =\columnwidth]{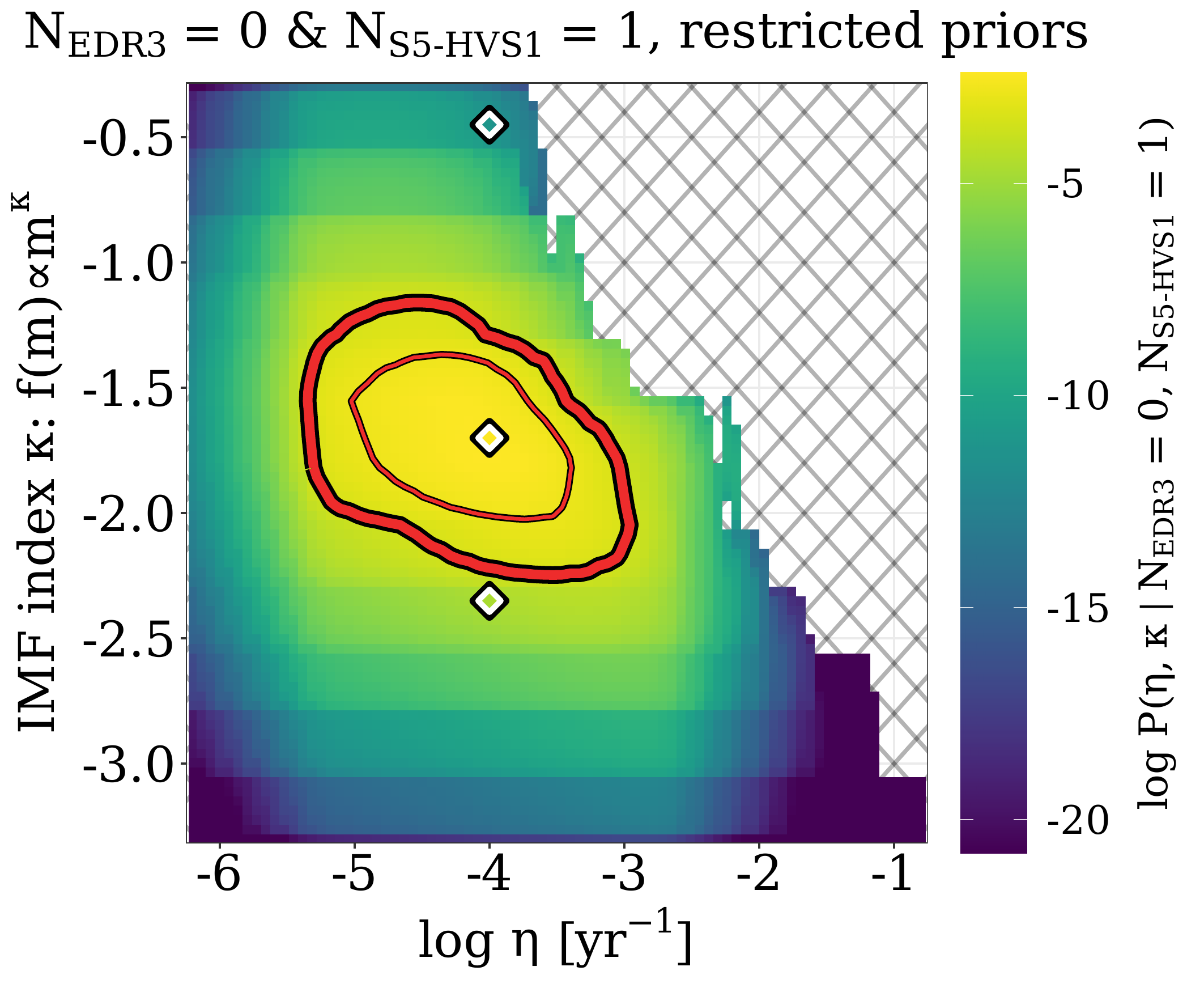}
    \caption{Colourscale shows (log) posterior probabilities for $\kappa-\eta$ model combinations when various data and priors are considered (see Sec. \ref{sec:bayes} \textcolor{black}{and panel titles}). The thin and thick red contours show the 68\% and 95\% Bayesian credible regions respectively. Black-and-white diamonds show our fiducial model $(\kappa=-1.7,\eta=10^{-4} \, \mathrm{yr^{-1}})$ as well as a model in which $\kappa=-2.3$ \citepalias{Salpeter1955}, and $\kappa=-0.45$ \citepalias{Bartko2010}. The hashed region shows models for which the posterior probability is zero.}
    \label{fig:constraints}
\end{figure*}

In Fig. \ref{fig:constraints} we show the outcome of our Bayesian modelling (Eq. \ref{eq:bayes}), broken up to show how each consideration of the data and priors impact the resulting posterior distributions. In the top-left panel,  the colourbar shows the posterior probabilities for $(\kappa, \eta)$ combinations if \textit{only} the lack of credible HVS candidates in the \textit{Gaia} EDR3 radial velocity catalogue is considered and and we assume uniform priors. The thin and thick red contours highlight the 68\% and 95\% Bayesian credible regions respectively. In this case these serve as upper limits, as any combination of sufficiently small $\eta$ and steep $\kappa$ is consistent with zero HVSs in EDR3. At our fiducial IMF slope, ejection rates in excess of $2\times10^{-4} \, \mathrm{yr^{-1}}$ can be excluded at >2$\sigma$. Models with an extremely top-heavy IMF such as that suggested by \citetalias{Bartko2010} can be discarded unless the HVS ejection rate is lower than $3\times 10^{-5} \, \mathrm{yr^{-1}}$. If the GC IMF is canonical \citepalias[$\kappa=-2.35$;][]{Salpeter1955}, ejection rates up to $10^{-3} \, \mathrm{yr^{-1}}$ are still allowed.

In the top-right panel of Fig. \ref{fig:constraints} we show posteriors if \textit{only} the existence of S5-HVS1 is considered and priors on $\kappa$ and $\eta$ are assumed uniform. This observation excludes regions of low $\eta$ / steep $\kappa$, as an S5-HVS1-like object is too rare an outcome from these models. Conversely, if the ejection rate is too large and the IMF too top-heavy, far more than one S5-HVS1 analogue is expected. The strip of models consistent with a single S5-HVS1-like object is degenerate in this space and includes our fiducial model within the 1$\sigma$ contour.

The bottom-left panel of Fig. \ref{fig:constraints} shows the joint posterior probabilities with uniform priors when both the lack of HVSs in \textit{Gaia} EDR3 and the existence of S5-HVS1 are considered. While no IMF slope can be excluded outright due to the degeneracy in this space, an HVS ejection rate above $2\times10^{-3} \, \mathrm{yr^{-1}}$ can be excluded unless the GC IMF is more top-light than a canonical \citetalias{Salpeter1955} IMF. A model in which the HVS ejection rate is $\eta=10^{-4} \, \mathrm{yr^{-1}}$ and the IMF is canonical can be excluded at $>1\sigma$ confidence, and a model in which $\eta=10^{-4} \, \mathrm{yr^{-1}}$ and $\kappa=-0.45$ \citepalias{Bartko2010} can be excluded at >2$\sigma$. Our fiducial $\kappa=-1.7$ is consistent with these observations for  $\eta=0.7_{-0.5}^{+1.5} \times10^{-4} \, \mathrm{yr^{-1}}$.

Finally, in the bottom-right panel of Fig. \ref{fig:constraints} we compute posterior distributions with when assuming our set of more restrictive priors. Together, the available HVS observational data and priors motivate a scenario in which $\kappa=-1.8_{-0.3}^{+0.4}$ and log $[\eta/\mathrm{yr^{-1}}] = -4.1_{-0.8}^{+0.6}$.  We point out that with these restrictive priors considered, our fiducial model $(\kappa=-1.7,\eta=10^{-4} \, \mathrm{yr^{-1}})$ sits comfortably within the 1$\sigma$ contour  and quite close to the maximum a posteriori model, though this is not surprising given the fact that our fiducial model was chosen in the first place based upon these priors. 

In summary, the constraints offered by EDR3 alone upon $\kappa$ and $\eta$ improve significantly upon those offered by \citetalias{Evans2022}, where only models in which $\eta\gtrsim3\times 10^{-2} \, \mathrm{yr^{-1}}$ could be excluded. By considering the existence of S5-HVS1 these constraints tighten further, particularly at low ejection rates and steep IMF slopes. There is no tension between these constraints and existing estimates of $\kappa$ and $\eta$ individually.

\subsection{Prospects for the future}

\begin{figure*}
    \centering
    \includegraphics[width =\columnwidth]{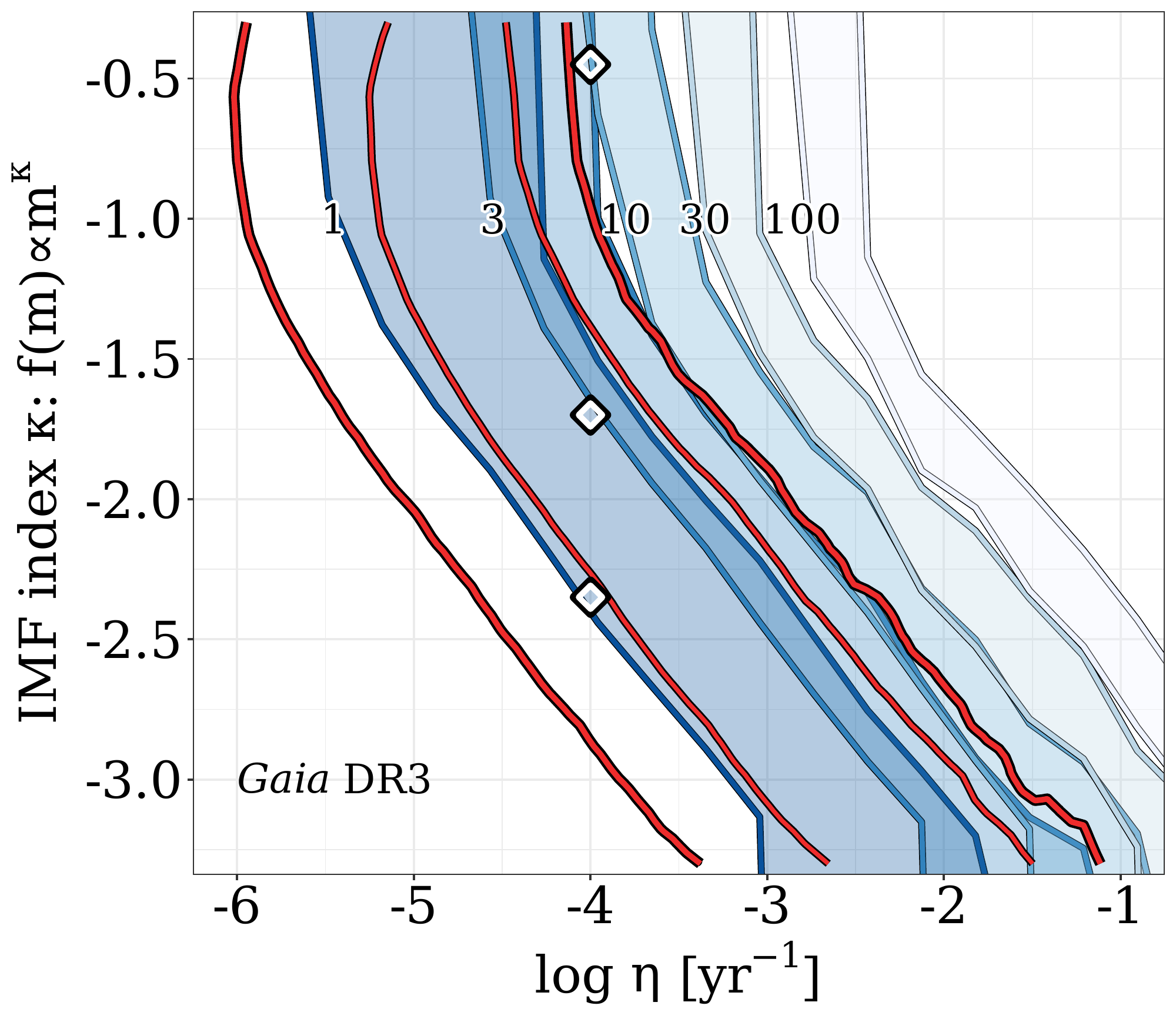}
    \includegraphics[width =\columnwidth]{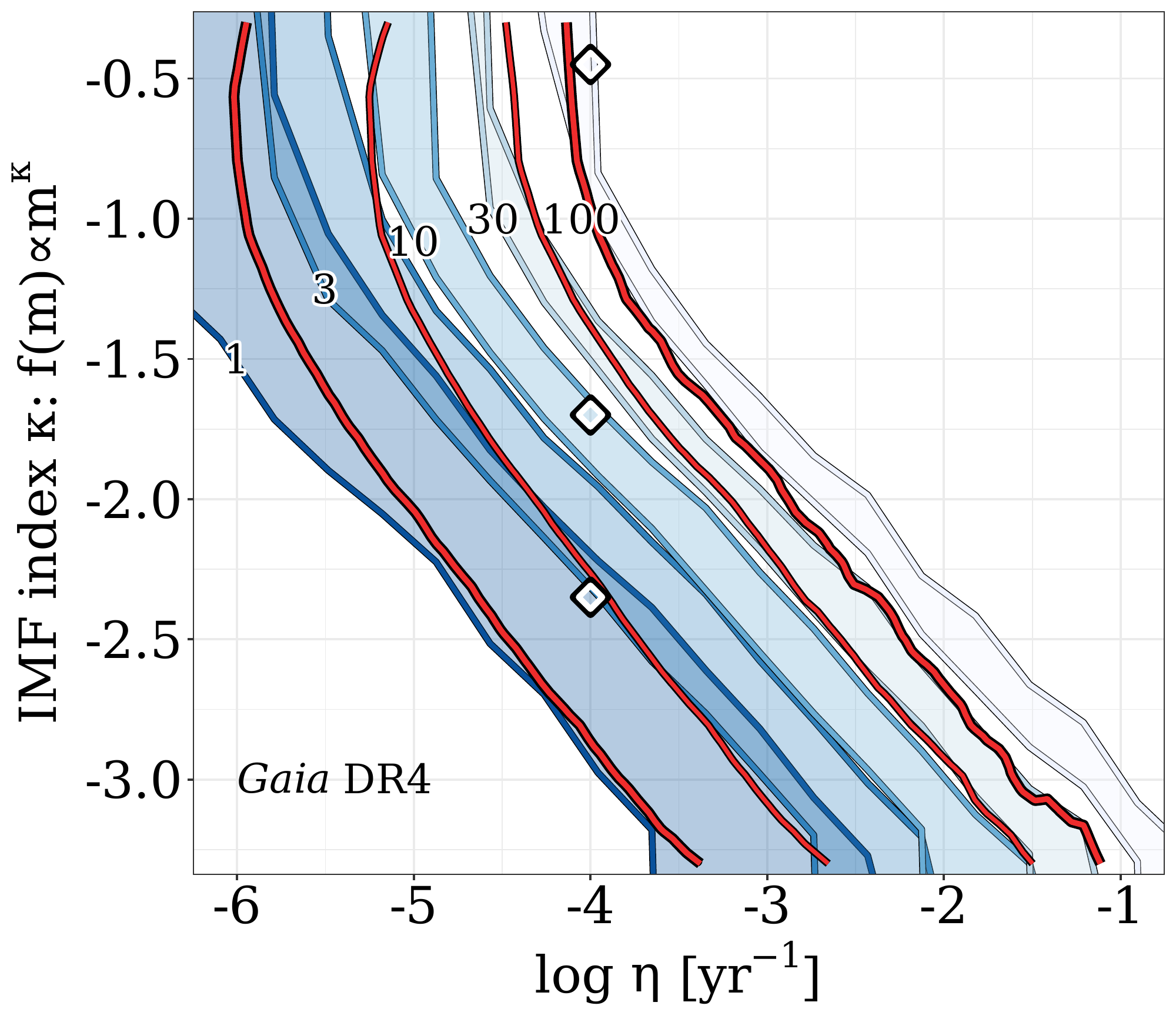}
    \caption{If 1/3/10/30/100 HVSs are discovered in \textit{Gaia} DR3 (left) and DR4 (right), the shaded coloured regions show the regions of the $\kappa-\eta$ parameter space consistent within 2$\sigma$ with these findings. The black-and-white diamonds indicate our fiducial ejection rate of $\eta=10^{-4} \, \mathrm{yr^{-1}}$ and fiducial $\kappa=-1.7$ \citepalias{Lu2013}, as well as $\kappa=-2.35$ \citepalias{Salpeter1955} and $\kappa=-0.45$ \citepalias{Bartko2010}. The red lines indicate the 68\% and 95\% credible regions for $\kappa-\eta$ as outlined in Sec. \ref{sec:bayes} when uniform priors are considered. } 
    \label{fig:HVSnumbers}
\end{figure*}

With the constraints outlined above, we are well-positioned to make specific predictions about the HVS population yet to be uncovered in \textit{Gaia} DR3 and DR4 and how these unearthed populations may improve constraints even further.

Each coloured band in Fig. \ref{fig:HVSnumbers} shows the \textcolor{black}{region of $\kappa-\eta$ space consistent with a specific number of HVSs appearing in \textit{Gaia} DR3 (left) and DR4 (right) at the 1$\sigma$ level}. For instance, if 100 high-confidence HVS are discovered in \textit{Gaia} DR3, the most-pale band shows the $\kappa-\eta$ models for which the $\pm1\sigma$ range of the predicted HVS population size includes 100. The red lines show the 68\% and 95\% credible intervals from our modelling constraints when uniform priors are considered on $\kappa$ and $\eta$ (Fig. \ref{fig:constraints}, lower left). Sampling from this posterior, we predict $0.8\pm0.7$ HVSs will be uncovered in DR3 and $4.9_{-3.7}^{+11.2}$ in DR4. Detecting HVS populations near these expectations will validate our methodology but may only offer modest improvements on model constraints. However, if zero or $\gtrsim$3 HVSs are detected in DR3 and/or $\lesssim$3 or $\gtrsim$20 HVSs are uncovered in DR4, updating posteriors with this new data will significantly change the maximum a posteriori model and tighten constraints considerably.

\begin{figure}
    \centering
    \includegraphics[width =\columnwidth]{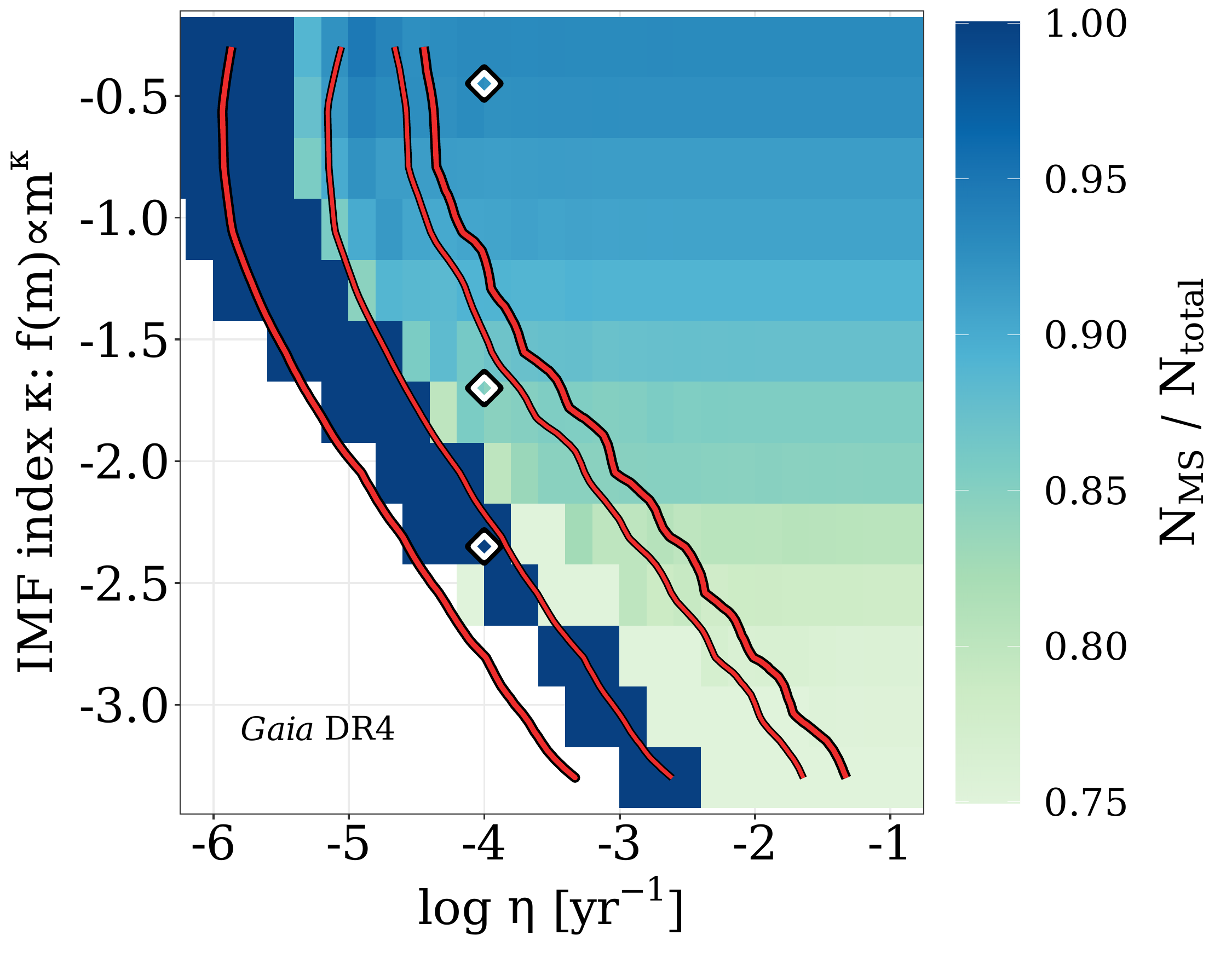}
    \caption{The \textcolor{black}{modal (most frequent) value for the} proportion of main sequence HVSs in \textit{Gaia} DR4. The red lines indicate the 68\% and 95\% credible regions in this parameter space (Fig. \ref{fig:constraints}, lower left). The \textcolor{black}{white region} shows models for which <1 total HVSs are expected.} 
    \label{fig:ratio}
\end{figure}

Our constraints are degenerate in $\kappa-\eta$ space: a larger ejection rate and steep IMF can predict the same number of HVSs as a lower ejection rate and shallower IMF. With \textit{Gaia} DR4 we can begin to break this degeneracy by examining the HVS population in greater detail. While HVSs ejected from the GC will on average be more massive if the IMF is top-heavy, the spread of HVS stellar masses is large and the detected DR4 HVS population is not likely to be numerous enough to provide insight into the IMF slope using the HVS mass distribution alone. More discriminating are the relative numbers of main sequence and evolved HVSs. We demonstrate this in Fig. \ref{fig:ratio}. Here we show how the \textcolor{black}{modal (most frequently occurring) proportion of main sequence stars among all DR4-detectable HVSs changes in $\kappa-\eta$ space, if $\geq1$ HVSs are expected at all. For the dark blue stripe of models towards lower ejection rates and steep IMFs, we predict only one or two HVSs in DR4 and they are most likely to be on the main sequence. For models in which more HVSs are predicted, the main sequence fraction of HVSs rises more or less monotonically} with increasing $\kappa$. If, for example, 12 HVSs in total are found in DR4 and 9 are on the main sequence, this would require and IMF no more top-heavy than $\kappa=-1.4$.

\section{Discussion}
\label{sec:discussion}

\subsection{Concerns, Caveats \& Alternative Assumptions}

The modelling and analysis presented here depends on a number of assumptions concerning the properties of stellar binaries in the GC, the mechanism of HVS ejection and the observational capabilities of \textit{Gaia}. In this subsection we comment on several of these assumptions and their impact on our results.

One assumption we have made here is that the Hills mechanism is solely responsible for HVS ejections from the GC -- the constraints on the GC stellar environment we present here apply exclusively to a Hills ejection scenario. While this is the most popular HVS ejection mechanism from the GC, alternative scenarios exist involving e.g. an as-of-yet undetected supermassive or intermediate-mass black hole companion to Sgr A* \citep[e.g.][]{Yu2003,Gualandris2005, Sesana2006, Sesana2007, Darbha2019, Rasskazov2019, Zheng2021}, a population of stellar mass black holes in the GC \citep{OLeary2008}, the disruption of infalling globular clusters \citep{Capuzzo2015, Fragione2016}\textcolor{black}{, or supernovae within GC binaries \citep{Zubovas2013, Bortolas2017, Hoang2022}}. Each of these mechanisms, if exclusively or partially responsible for HVSs, would warrant a different approach to the ejection model.

In predicting the future \textit{Gaia} HVS population, we have assumed that astrometric and spectroscopic solutions will have zero systematic error (see discussion in \citetalias{Evans2022}) and that the detected HVS population will not be contaminated by fast stars ejected from outside the GC. In \citetalias{Evans2022} Appendix B we show that the latter is not a pressing concern -- 90 to 95 per cent of HVSs detectable by \textit{Gaia} will have trajectories which unambiguously suggest an origin in the GC. 

In this work we have assumed that the mass ratios ($q$) among GC binaries follow a power-law distribution, i.e. $f(q) \propto q^{\gamma}$, where in each iteration $\gamma$ is drawn at random in the interval [-2,+2]. A feature we have not allowed in this distribution is the so-called ``twin" phenomenon -- the observed statistical excess of equal or nearly equal-mass binary systems across a range of total mass and orbital separation ($q\gtrsim$0.95) \citep{Lucy1979, Tokovinin2000, Moe2017, ElBadry2019}. We test the impact of this feature by running another suite of simulations in the extreme case in which \textit{all} binaries in the GC are equal-mass. While our predictions for \textit{Gaia} DR4 change slightly ($18.5_{-4.4}^{+4.6}$ total HVSs in our fiducial model compared to $10.9_{-4.2}^{+4.8}$ in our original prescription), the constraints on the GC IMF and HVS ejection rate from \textit{Gaia} EDR3 and S5-HVS1 remain largely unchanged.

We assess the impact of the IMF functional form with a similar test. We have assumed a single power-law in this work ($f(m)\propto m^{\kappa}$), while other canonical IMFs have broken power law \citep{Kroupa2001} or log-normal \citep{Chabrier2003} forms. Modern infrared observations of the GC \citepalias{Bartko2010, Lu2013} are not sensitive to low-mass stars, so they cannot constrain a change in the IMF slope in the subsolar regime. We run yet another suite of simulations where we vary the IMF slope only in the $m\geq0.5 \, \mathrm{M_\odot}$ regime and the IMF log-slope remains fixed at $-1.3$ in the range $0.08 \, \mathrm{M_\odot} \leq m<0.5 \, \mathrm{M_\odot}$ \citep{Kroupa2001}. Constraints from \textit{Gaia} EDR3 and S5-HVS1 become slightly more vertical in $\kappa-\eta$ space as we now expect more (fewer) HVSs for $\kappa<-1.3$ ($\kappa>-1.3$) when compared to a single power-law, but otherwise remain unchanged. For our fiducial model, this prescription would predict $14.2_{-5.0}^{+5.8}$ total HVSs in \textit{Gaia} DR4.

\textcolor{black}{Throughout this work we have assumed a constant HVS ejection rate and, implicitly, a constant star formation rate in the GC. In our fiducial model ($\kappa=-1.7$, $\eta=10^{-4} \, \mathrm{yr^{-1}}$), only stars with $t_{\rm flight} \lesssim 20 \, \mathrm{Myr}$ would be bright enough to appear in any \textit{Gaia} data release with precise astrometry and a measured radial velocity, and only HVSs with $t_{\rm flight} \lesssim65 \, \mathrm{Myr}$ would be bright enough to appear in S$^5$ as an S5-HVS1 analogue. Our constraints on $\eta$ in this work can be therefore thought of as applying only to the typical ejection rate over the last few tens of Myr, since we base these constraints only on (un-)observed HVSs ejected in the relatively recent past. By a similar token, only the GC star formation history within the last $\sim$0.5 Gyr is relevant for this work -- HVSs older than this would not be detectable in both \textit{Gaia} EDR3 and S$^5$. While there is evidence suggesting that the star formation rate in the GC has been non-continuous throughout the history of the Galaxy and has in fact increased slightly within the last $\sim$100 Myr \citep{Pfuhl2011, NoguerasLara2020}, a constant star formation rate within the last 0.5 Gyr is a reasonable assumption \citep[see also][]{Figer2004}.}

\textcolor{black}{Stars age in our model according to standard single stellar evolution prescriptions. While binary interactions can be ignored since we have required that HVS progenitor binaries remain sufficiently well-separated, the extremity of the GC environment may still influence stellar evolution. Of particular interest is nuclear activity from Sgr A*. $\sim$40 per cent of \textit{Gaia} EDR3-detectable HVSs in our fiducial model were ejected within the last 8 Myr, and evidence for a Seyfert-level flare from the GC $\sim$2-8 Myr ago has been mounting in recent years \citep[see][and references therein]{BlandHawthorn2019, Cecil2021}. Active galactic nuclei (AGN) are known to impact the evolution of stars within them -- accretion from the AGN gas disc can extend the hydrogen-burning lifetime of low-mass stars and increase their total mass \citep{Cantiello2021, Dittmann2021, Jermyn2022}. If prior episodes of nuclear activity in the GC have impacted a non-negligible fraction of HVS progenitor binaries, then their evolutionary states and apparent magnitudes may be inaccurately estimated.}

\textcolor{black}{One assumption in our model is that HVSs are equally likely to be ejected at any point during their lifetime. This is motivated by the fact that existing HVS candidates do not appear to be biased towards particular ages. We also assume that the IMF among the primaries of HVS progenitor binaries matches the IMF of stars in the GC region as a whole, and that this IMF remains constant in time. These assumptions mean that some massive \textit{Gaia}-detectable mock HVSs in our simulations must be ejected quite shortly after formation. Among (E)DR3-visible HVSs, the typical age of an HVS at the moment of ejection is 100 Myr. Among DR4-visible HVSs, however, these median age at ejection drops to only $\sim$10 Myr. Theoretical works indicate that diffusing a binary into the Sgr A* loss cone to soon after formation may be problematic \citep{Yu2003, Wang2004, Merritt2004, Perets2007}. If an unrealistic number of young HVSs are being ejected in our model, predictions for the \textit{Gaia} DR4 HVS population may be overestimated.}

\subsection{Previous work and other HVS (non-)detections}

Prior works have used HVS non-detections to constrain the GC ejection rate. \citet{Kollmeier2009} infer an ejection rate for HVSs of spectral type F and G of $\eta_{\rm F}<6\times10^{-5} \, \mathrm{yr^{-1}}$ and $\eta_{\rm G}<3\times10^{-4} \, \mathrm{yr^{-1}}$ respectively upon finding zero old, unbound HVS candidates among stars with measured radial velocities in the Sloan Digital Sky Survey \citep[SDSS;][]{York2000}. Our results are consistent with these constraints -- at our fiducial $\kappa$, our constraints in the lower left panel of Fig. \ref{fig:constraints}  indicate $\eta_{\rm F}\lesssim8\times10^{-5} \, \mathrm{yr^{-1}}$ at 2$\sigma$ confidence and $\eta_{\rm G}\lesssim6\times10^{-5} \mathrm{yr^{-1}}$. \citet{Kollmeier2010} similarly find zero metal-rich old HVSs in SEGUE-2 \citep{Yanny2009}. They deduce that the ejection rate of $5 \, \mathrm{Gyr}$-old, solar-metallicity HVSs which reach a Galactocentric velocity of $500 \, \mathrm{km \ s^{-1}}$ at the Solar circle is $<4.1\times10^{-4} \, \mathrm{yr^{-1}}$ per logarithmic unit of stellar mass, again consistent with this work. Notably, \citet{Kollmeier2010} also conclude that the GC ejects $\sim$5.5 times as many F/G stars as B stars, corresponding to a quite top-heavy GC IMF ($\kappa\approx-0.6$).

In principle, HVS null detections (to date) in other ground-based surveys such as RAVE \citep{Steinmetz2006}, LAMOST \citep{Zhao2012}, GALAH \citep{DeSilva2015}, APOGEE \citep{Majewski2017} and H3 \citep{Conroy2019} could also be used to place constraints on the GC stellar environment. Properly considering HVS non-detections in all these surveys combined would require careful modelling of each individual survey's selection function and observational systematics, with no guarantee that constraints would improve relative to those provided by \textit{Gaia} alone. The advantage of \textit{Gaia} is its coverage, its catalogue size, its (relatively) well-modelled spectroscopic selection function and its ability to measure 3D velocities without complementary observations from other surveys -- we focus on it here and defer a holistic treatment of all available Galactic surveys to future work. 

Another option is to consider the HVS candidates in the MMT HVS Survey \citep{Brown2009, Brown2014}, which uncovered tens of HVS candidates by targeting $[2.5, 4] \, \mathrm{M_\odot}$ stars in SDSS for follow-up spectroscopic observation. While it is relatively straightforward to select analogues for these candidates from among our mock HVS samples using the SDSS footprint and mock SDSS photometry, it is unclear how many and which MMT HVS candidates are genuine GC-ejected HVSs. A GC origin remains plausible for a $\sim$dozen candidates in the MMT HVS Survey \citep[see][]{Brown2018, Irrgang2018, Kreuzer2020}, however, the distribution of possible ejection locations for many candidates is many times larger than the entire Galactic disc. While valuable analyses can be done assuming all of these candidates are genuine HVSs \citep[e.g.][]{Kollmeier2009, Kollmeier2010, Rossi2017}, due to this ambiguity we opt not to consider the MMT HVS Survey when testing our model predictions.

\begin{figure*}
    \centering
    \includegraphics[width =2\columnwidth]{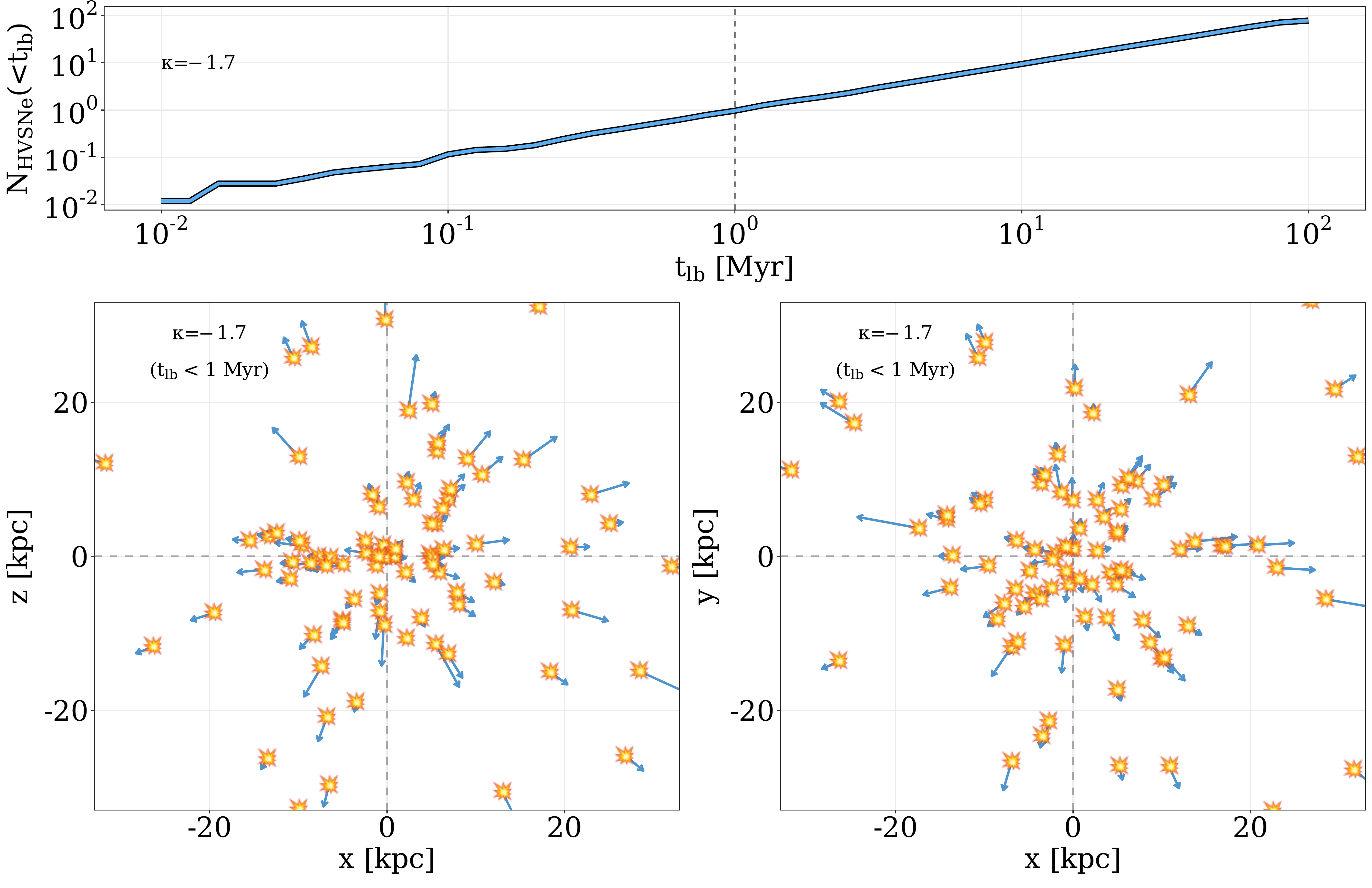}
    \caption{\textit{Top:} Cumulative distribution of hypervelocity supernovae (HVSNe) lookback times $t_{\rm lb}$ in our fiducial model. Dashed vertical line indicates $t_{\rm lb}=1 \, \mathrm{Myr}$ ago. \textit{Bottom}: The locations of HVSNe occurring within the last Myr in the Galactocentric cartesian x-z (left) and x-y (right) planes, stacked over 50 runs. Blue vectors indicate the velocity of the progenitor HVS at the moment of core collapse.}
    \label{fig:SNe}
\end{figure*}

\subsection{Hypervelocity curios} \label{sec:discussion:curios}

\subsubsection{Hypervelocity standard candles}

A keen-eyed reader may notice that a minority population of HVS candidates in \textit{Gaia} DR3 and DR4 are \textcolor{black}{core helium-burning stars currently on a blue loop phase of evolution (see Fig. \ref{fig:HR})}. Many such stars will cross the so-called instability strip; stars in this region of the Hertzsprung-Russell diagram are unstable to radial oscillations and may stand out as classical Cepheid variable stars.

Due to the correlation between their pulsational periods and intrinsic luminosities \citep{Leavitt1912}, heliocentric distances to \textit{Gaia} Cepheids can currently be determined to a precision of a few per cent \citep[see][]{Owens2022}. With such precise distance estimations, the birthplaces and Galactocentric velocities of hypervelocity Cepheid (HVC) candidates can be tightly constrained. HVCs appearing in the \textit{Gaia} radial velocity catalogues would therefore \textit{not} need to satisfy our strict $20\%$ relative parallax error cut to be identified as a high-quality HVS candidate. Furthermore, a radial velocity measurement may not even be necessary for HVCs with large tangential velocities: with a precise distance estimate, uncertainties on tangential velocities will be small and the HVS candidate's origin can be well-constrained even in the absence of full 3D velocity information. In our fiducial model we predict $1.9_{-1.0}^{+1.2}$ HVCs to appear in the DR4 source catalogue ($G<20.7$). Of these, however, only $1.1_{-0.9}^{+1.2}$ will be bright enough to appear in the radial velocity catalogue. Therefore, while prospects are not particularly promising, with some luck DR4 may supply the first known GC-ejected hypervelocity standard candle. 

We note as well that a significant fraction of HVSs detectable in \textit{Gaia} DR3 and DR4 reside in the so-called `red clump' at $T_{\rm eff}\sim5000 \, \mathrm{K}$ and $L\sim100 \, \mathrm{L_\odot}$ (Fig. \ref{fig:HR}), corresponding to low-mass core helium-burning stars \citep[see][for a review]{Girardi2016}. With its roughly fixed absolute magnitude, the red clump can be used as a standard candle to measure distances \citep{Cannon1970}. Since \textit{Gaia} HVSs will be located all across the sky in regions of differing extinction, calibrating a clean red clump HVS sample using \textit{Gaia} optical photometry alone is unfeasible. This does, however, support the attractive possibility of searching for red clump HVSs in cross-matched combinations of \textit{Gaia} with infrared Galactic surveys \citep[see][]{Luna2019}.

\subsubsection{Hypervelocity supernovae}

Within this work we have shown that there exists a population of HVSs in the Galaxy which are at late stages of stellar evolution. It is natural, then, to wonder about the deaths of these HVSs. Massive ($m\gtrsim8 \, \mathrm{M_\odot}$) HVSs may undergo core-collapse supernovae (CCSNe) whose explosion or remnant could be detected. It is important to note, however, that not all core-collapse events can be associated with a supernova explosion. A significant fraction may collapse directly to a black hole without a luminous electromagnetic signature \citep{Kochanek2008}. The precise outcome for a $m_{\rm ZAMS} \gtrsim 14 \, \mathrm{M_\odot}$ core-collapse event depends intimately on subtle aspects of its progenitor star's structure in its final moments -- a simple mapping between progenitor and outcome does not exist \citep{OConnor2011,Pejcha2016, Ertl2016, Sukhbold2016} and robustly modelling this is beyond the scope of this work. Regardless, with some simple assumptions we can use our simulation framework to explore the occurrence of hypervelocity supernovae (HVSNe) in the Galaxy.

Since stars which do not survive until the present day are removed in our methodology as described in Section \ref{sec:methods}, we perform another suite of simulations. Using the same $\kappa-\eta$ model grid, we eject and propagate only $m>8 \, \mathrm{M_\odot}$ stars which were main sequence or evolved stars at time of ejection, but are compact remnants today according to our stellar evolution prescription. We make the simple assumption that $m_{\rm ZAMS}\geq20 \, \mathrm{M_\odot}$ stars tend to implode rather than go supernova \citep[see][]{Ertl2016, Sukhbold2018, Sukhbold2020}, and we remove them for the sample. A star evolves as it is propagated through the Galactic potential, and we end the orbital integration at the first timestep in which the star is a compact remnant. The star is assumed to undergo a CCSNe at this time and we record the location in the Galaxy and the lookback time $t_{\rm lb}$ ago at which this happened.

We show the results of this investigation in Fig. \ref{fig:SNe}. The top panel shows the cumulative distribution of $t_{\rm lb}$ over the last $100 \, \mathrm{Myr}$. Our fiducial model predicts that $75_{-23}^{+31}$ core-collapse HVSNe have occurred during this time period, with only $1.0_{-0.8}^{+1.1}$ occurring within the last Myr. Assuming the rate of CCSNe in the Milky Way is $\sim[1-2] \times 10^{-2} \, \mathrm{yr^{-1}}$ \citep[see][and references therein]{Rozwadowska2021}, HVSNe then represent $\sim0.005-0.01$ per cent of Galactic CCSNe. In the bottom panels of Fig. \ref{fig:SNe} we show the locations of $t_{\rm lb}<1 \, \mathrm{Myr}$ in the Galaxy in Galactocentric cartesian x-z plane (left) and x-y plane (right), stacked over 100 iterations. Blue arrows indicate the Galactocentric velocity of the HVSNe progenitor star at the moment of core collapse. While most HVSNe occur in the inner few kpc of the Galaxy arising from short-lived massive stars, nearly half will be offset from the Galactic disc by $10 \, \mathrm{kpc}$ or more. These events would be characterized by their exceptional line of sight: 7 in 10 HVSNe would satisfy $|v_{\rm rad}|\geq800 \, \mathrm{km \ s^{-1}}$.

While Galactic HVSNe are quite rare, they invite the prospect of observing HVSNe in other galaxies.  New and ongoing transient surveys such as the Zwicky Transient Facility Bright Transient Survey \citep{Fremling2020, Perley2020}, the All Sky Automated Survey for Supernovae \citep{Shappee2014} and the Asteroid Terrestrial-impact Last Alert System \citep{Tonry2018} scan the optical sky nearly nightly and to date have observed $\sim$hundreds of extragalactic CCSNe in the local ($z\lesssim0.1$) Universe. With the Rubin Observatory's upcoming Legacy Survey of Space and Time \citep{Ivezic2019}, this rate of CCSNe detections is expected to increase tenfold. Extragalactic HVSNe might be uncovered in such surveys by searching for events significantly offset from the disc of their host and/or with significant peculiar velocities with respect to their host. \textcolor{black}{The initial mass functions and HVS ejection rates within the nuclei of external galaxies will depend on their accretion history, star formation history, and history of nuclear activity}. Such HVSNe observations would be the first observational evidence of hypervelocity ejections outside the Local Group and would join tidal disruption event rate observations \citep[see][]{Bortolas2022} as a valuable tool for directly investigating the nuclei of galaxies in the local Universe.

\section{Conclusions}
\label{sec:conclusions}

In this work we simulate the ejection of hypervelocity stars (HVSs) from the Galactic Centre (GC) via the tidal breakup of stellar binaries following dynamical encounters with Sgr A*. We expand upon the previous work of \citet{Evans2022} by investigating evolved HVSs as well as main sequence ones, as these evolved HVSs would more easily appear in current data releases from the European Space Agency's \textit{Gaia} mission. By considering that lack of known HVSs in \textit{Gaia} EDR3 with precise astrometry and radial velocities as well as the existence of the HVS candidate S5-HVS1 \citep{Koposov2020}, we place robust and competitive constraints on the stellar initial mass function (IMF) \textcolor{black}{among the primaries of HVS progenitor binaries and the} ejection rate of HVSs from the GC. Using these constraints, we make predictions about the evolved and main sequence HVS populations to appear in upcoming \textit{Gaia} data releases.

Our findings can be summarized as follows:

\begin{itemize}
    \item For a fiducial model in which the IMF is a single power law ($f(m)\propto m^{\kappa}$) with $\kappa$ = $-1.7$ \citep{Lu2013} and the HVS ejection rate $\eta$ is $10^{-4} \, \mathrm{yr^{-1}}$ \citep[see][]{Brown2015rev}, $<1$ high-confidence HVSs \textit{in total} are expected in the radial velocity catalogues of \textit{Gaia} DR2 and EDR3. This is consistent with observations \citep[e.g.][]{Marchetti2018, Hattori2018, Marchetti2021} (Fig. \ref{fig:NHVSkappa}).
    \item Predicted numbers of observed HVSs in all \textit{Gaia} data releases are degenerate in the IMF index-ejection rate parameter space (Fig.~\ref{fig:HVSkappaeta} and following Figures).
    \item For $\kappa=-1.7$, the lack of GC-ejected HVS candidates in \textit{Gaia} EDR3 disfavours  ejection rates above $2\times10^{-4} \, \mathrm{yr^{-1}}$. Larger ejection rates are only allowed if the IMF is top-light (upper left panel of Fig. \ref{fig:constraints}). 
    \item Accounting as well for the existence of the S5-HVS1, we obtain tighter constraints that additionally exclude the {\it low} ejection rate - bottom-light IMF region of the parameter space (lower left panel of Fig.~\ref{fig:constraints}). At $\kappa=-1.7$ the evidence favours an HVS ejection rate of $\eta=0.7_{-0.5}^{+1.5} \times10^{-4} \, \mathrm{yr^{-1}}$. \textcolor{black}This favoured ejection rate grows (shrinks) as the IMF becomes more bottom-heavy (top-heavy).
    \item Our derived constraints predict $0.8\pm0.7$ HVSs to be present in the \textit{Gaia} DR3 radial velocity catalogue with precise astrometry, and $4.9_{-3.7}^{+11.2}$ in \textit{Gaia} DR4. The majority of DR3 HVSs will be core helium-burning, while main sequence HVSs will dominate in DR4 (Figs. \ref{fig:NHVSkappa} and \ref{fig:HVSnumbers}).
\end{itemize}

With this work and with \citetalias{Evans2022}, we have shown that where HVSs \textit{are not} is equally as interesting as where they are. We have demonstrated in this work that evolved HVSs in the context of \textit{Gaia} are a powerful tool for constraining the GC environment. This work shows that competitive constraints on the stellar initial mass function and HVS ejection rate in the GC can be gleaned from a small number of HVS (non-)detections. With future \textit{Gaia} data releases and with complementary upcoming Galactic spectroscopic surveys such as WEAVE \citep{Dalton2012} and 4MOST \citep{deJong2019}, HVS observations will gain even more prominence as an avenue for studying the supermassive black hole at the centre of our Galaxy and its interactions with its environment.

\section*{Acknowledgements}

\textcolor{black}{The authors thank the anonymous referee for their valuable feedback.} They also wish to Anthony Brown and Rainier Sch\"{o}del for helpful comments and Sergey Koposov, Warren Brown, Sill Verberne and Alonso Luna for enlightening conversations. FAE acknowledges funding support from the Natural Sciences and Engineering Research Council of Canada (NSERC) Postgraduate Scholarship. TM acknowledges an ESO fellowship. 
EMR acknowledges that this project has received funding from the European Research Council (ERC) under the European Union’s Horizon 2020 research and innovation programme (Grant agreement No. 101002511 - VEGA P).

\section*{Data Availability}

The simulation outputs underpinning this work can be shared upon reasonable request to the corresponding author. \textcolor{black}{These simulations were produced using the \texttt{speedystar} package, publicly available at \url{https://github.com/fraserevans/speedystar}}.


\bibliographystyle{mnras}
\bibliography{HRS}

\bsp	
\label{lastpage}
\end{document}